\documentclass[fp,twocolumn]{jpsj3}

\usepackage{cite}

\usepackage{amsmath}
\usepackage{txfonts}
\usepackage{bm}
\usepackage{color}

\title{
Possible superconductivity induced by a large spin-orbit coupling in carrier doped iridium oxide insulators: A weak coupling approach
}

\author{
Kazutaka Nishiguchi$^{1,2}$\thanks{nishiguchi@pearl.kobe-u.ac.jp}, 
Tomonori Shirakawa$^{1,3,4,5}$, 
Hiroshi Watanabe$^{5,6}$, 
Ryotaro Arita$^{7,8}$, 
and Seiji Yunoki$^{1,4,5}$
} 

\inst{
$^1$Computational Condensed Matter Physics Laboratory, RIKEN Cluster for Pioneering Research (CPR), Saitama 351-0198, Japan \\
$^2$Graduate School of Science Technology, and Innovation, Kobe University, Nada-ku, Kobe 657-8501, Japan \\
$^3$International School for Advanced Studies (SISSA), Via Bonomea 265, 34136, Trieste, Italy \\
$^4$Computational Materials Science Research Team, RIKEN Center for Computational Science (R-CCS),  Hyogo 650-0047,  Japan \\
$^5$Computational Quantum Matter Research Team, RIKEN, Center for Emergent Matter Science (CEMS), Saitama 351-0198, Japan \\
$^6$ Waseda Institute for Advanced Study, Waseda University, Shinjuku, Tokyo 169-8050, Japan \\
$^7$Department of Applied Physics, University of Tokyo, Tokyo 113-8656, Japan\\
$^8$First-Principles Materials Science Research Team, RIKEN, Center for Emergent Matter Science (CEMS), Saitama 351-0198, Japan
} 

\abst{
We study possible superconductivity in a carrier-doped iridium oxide insulator Sr$_{2}$IrO$_{4}$ 
based on  an effective $t_{2g}$ three-orbital Hubbard model on the square lattice with a large spin-orbit coupling (SOC). 
Numerically solving the linearized Eliashberg equation for the superconducting (SC) gap function with the random phase approximation, 
we systematically examine both singlet and triplet SC gap functions with possible pairing symmetries and the parameter dependence of the superconductivity.
For the realistic SOC $\lambda$ and Hund's coupling $J/U$ relevant to Sr$_{2}$IrO$_{4}$, namely, for a large $\lambda$ and small $J/U$ region, 
we find that 
the intra-band antiferromagnetic (AF) pseudospin $\bm{j}_{\text{eff}} = -\bm{l}+\bm{s}$ fluctuations favor 
a $d_{x^{2}-y^{2}}$-wave pseudospin $j_{\text{eff}}=1/2$ singlet pairing in the electron-doping. 
We also find that 
the $d_{x^{2}-y^{2}}$-wave pairing is more stabilized with increasing the SOC and decreasing the Hund's coupling. 
Furthermore, we show for a small $\lambda$ and large $J/U$ region that 
an $s_{\pm}$-wave singlet pairing is favored in the hole-doped region. 
The origin of the $s_{\pm}$-wave pairing is due to the inter-band pair scattering 
arising from the intra-orbital AF spin $\bm s$ fluctuations. 
Although the possibility of a pseudospin triplet pairing is considered, 
we find it always unfavorable for all parameters studied here. 
The experimental consequences for other strongly correlated materials with a large SOC are also discussed. 
} 

\newcommand{\bra}  {\langle}
\newcommand{\ket}  {\rangle}

\begin{document}

\maketitle

\section{Introduction} 

Superconductivity is one of the most fundamental and intriguing quantum mechanical phenomena in condensed matter physics, 
and the discovery of new exotic superconductors 
such as high-$T_{\text{c}}$ cuprate superconductors,~\cite{BednorzMuller86} organic superconductors,~\cite{Jerome80}  and iron-based superconductors~\cite{Kamihara06} 
has stimulated extensive experimental as well as theoretical studies.~\cite{Uemura09, Scalapino12, Aoki12} 
The origins of these exotic superconductivity have been still under active debate. 
However, the current consensus is that 
the high-$T_{\text{c}}$ cuprate superconductor is an example of strongly correlated systems, 
and the antiferromagnetic (AF) fluctuations favor the unconventional $d$-wave superconducting (SC) state 
once mobile carriers are introduced into the AF Mott insulating phase. 
The highest critical temperature ($T_{\text{c}}$) at ambient pressure to date is achieved by hole doping in high-$T_{\text{c}}$ cuprates.~\cite{Leggett06b} 

Recently, 5$d$ transition metal oxide Sr$_{2}$IrO$_{4}$ and other related iridium oxides have attracted much interests 
both experimentally~\cite{Kini_2006, Kim08, Kim09, Franke11, Ishii11, Kim12, Fujiyama12, Lee12, Hsieh12, Li13, Wang13, Yamasaki14, Dai14, Uchida14, Moser14, Cui16, Yamasaki16, Souri17} 
and theoretically~\cite{Jackeli09, Jin09, Watanabe10, Martins11, Arita12, BHKim12, Carter12, Katukuri12, Igarashi13, Watanabe14, Lovesey14, Igarashi14, Sato15, Wang15, Liu15, BHKim16, Andrzej16, Arakawa16, AJKim17, Zhou17, Mohapatra17, Martins_2017, Moutenet18, Bhandari18, Martins18, Liu18, Sato19}  
because of the unexpected properties and novel phenomena arising from highly entangled spin and orbital degrees of freedom 
due to the large relativistic spin-orbit coupling (SOC)~\cite{Cao_2018}. 
Sr$_{2}$IrO$_4$ is in a layered perovskite structure forming a two-dimensional square lattice composed of Ir atoms~\cite{Randall57,Huang94}, 
and thus it is similar to a parent compound of high-$T_{\text{c}}$ cuprate superconductors such as La$_{2}$CuO$_{4}$~\cite{Pickett89}. 
The five 5$d$ electrons of Ir ion occupy the $t_{2g}$ orbitals, which are energetically lower than the $e_g$ orbitals due to the crystal field. 
The large SOC lifts the sixfold degenerate $t_{2g}$ orbitals (including spin degree of freedom) 
into the fourfold degenerate effective total angular momentum $j_{\text{eff}}=|-\bm{l}+\bm{s}|=3/2$ states and the doubly degenerate $j_{\text{eff}}=1/2$ states; 
Four electrons fill the $j_{\text{eff}}=3/2$ states and the remaining electron occupies one of the $j_{\text{eff}}=1/2$ states. 

Several experiments have revealed that 
the states around the Fermi level is contributed mostly by these $j_{\text{eff}}=1/2$ states~\cite{Kim09, Wang13, Uchida14, Yamasaki16}, 
which are antiferromagnetically ordered in the two-dimensional plane at low temperatures with weak ferromagnetic moment~\cite{Crawford94,Cava94,Shimura95,Cao98}. 
Furthermore, the antiferromagnetic exchange has been estimated to be as large as 50--100 meV~\cite{Kim08, Kim09, Kim12, Fujiyama12}. 
Therefore, there are many similarities between La$_{2}$CuO$_{4}$ and Sr$_{2}$IrO$_4$, as summarized in Table~\ref{tab:comp}. 
Because of these similarities, the possibility of superconductivity in the Ir oxides has been proposed theoretically 
once mobile carriers are introduced into the $j_{\text{eff}}=1/2$ antiferromagnetic insulator~\cite{Wang11, Watanabe13, Meng14, Yang14, Sumita17}. 
Although the superconductivity has not been experimentally observed yet in these Ir oxides~\cite{Okabe13, Chen15, Gretarsson16, Ito16, Chen16, Cheng16, Han16, Horigane18}, 
anomalous normal state properties similar to those observed often for under doped cuprates have been reported~\cite{Kim14, Torre15, Yan15, Liu15_2, Zhao16, Kim16, Battisti17, Seo17, Terashima17, Chen18}. 

\begin{table} 
\begin{center} 
\begin{tabular}{c||c|c}
    \hline \hline
    {} & La$_{2}$CuO$_{4}$~\cite{Scalapino12, Uemura09, Aoki12, Leggett06b} & Sr$_{2}$IrO$_4$\cite{Randall57,Huang94,Kim08, Kim09, Kim12, Fujiyama12, Cao_2018} \\
    \hline
    Crystal structure & K$_{2}$NiF$_{4}$ type &K$_{2}$NiF$_{4}$ type \\
    Number of holes  &one hole per $d_{x^2-y^2}$ & one hole per $j_{\rm eff}=1/2$ \\
    Ground state &  $s=1/2$ AF insulator & $j_{\rm eff}=1/2$ AF insulator\\
    Neel temperature &  $T_{\rm N}=325$ K & $T_{\rm N}=230$ K \\
    Magnetic exchange &  $J_{\rm ex}\approx125$ meV & $J_{\rm ex}\approx50-100$ meV \\
    Carrier doping & high-$T_{\text{c}}$ superconductivity & ?? \\
    \hline \hline 
\end{tabular} 
\caption{
Comparison between $3d$ transition metal oxide La$_{2}$CuO$_{4}$ and 5$d$ transition metal oxide Sr$_{2}$IrO$_4$. 
}
\label{tab:comp} 
\end{center} 
\end{table}

In the previous theoretical studies, various numerical techniques 
such as variational Monte Carlo method, functional renormalization group method, and dynamical mean field theory with continuous time quantum Monte Carlo method 
are employed to show that a $d_{x^{2}-y^{2}}$-wave pairing is favored in the electron-doped region\cite{Wang11, Watanabe13} 
and a $s_{\pm}$-wave\cite{Yang14} or $p$-wave\cite{Meng14} pairing in the hole-doped region. 
However, in these studies, the parameter dependence of these SC states has not been thoroughly explored. 
This is precisely one of our purposes in this paper. 

To investigate possible superconductivity in carrier doped 5$d$ transition metal oxide Sr$_{2}$IrO$_{4}$, 
we consider a $t_{2g}$ three-orbital Hubbard model with the SOC term on the square lattice as an effective model for Sr$_{2}$IrO$_{4}$.  
For this end, the linearized Eliashberg equation for the SC gap function is solved with the random phase approximation (RPA). 
To take account of the SOC in the RPA framework, 
we formulate the spin-dependent and multi-orbital RPA which can apply to a multi-orbital Hubbard model with the SOC. 
With this formulation, 
we discuss the possible pairing symmetries of the SC gap functions and their pairing mechanisms arising from the fluctuations based on the weak-coupling theory. 
The stability of the possible SC states is systematically studied by varying the model parameters such as the SOC $\lambda$, carrier concentration, and Hund's coupling $J/U$. 

Our weak-coupling analysis reveal that 
that a $d_{x^{2}-y^{2}}$-wave pseudospin $j_{\text{eff}}=1/2$ singlet pairing is favored for the parameters relevant to Sr$_2$IrO$_4$, namely, for large $\lambda$ and small $J/U$. 
We find that the $d_{x^{2}-y^{2}}$-wave pseudospin $j_{\text{eff}}=1/2$ singlet pairing is induced by the intra-band AF pseudospin $\bm{j}_{\text{eff}}$ fluctuations, 
and that the $d_{x^{2}-y^{2}}$-wave pairing is stabilized by the large SOC especially in the electron-doped region.
In contrast, we find that a $s_{\pm}$-wave pseudospin singlet pairing is favored in the hole-doped region when the SOC is small and $J/U$ is large. 
The origin of the $s_{\pm}$-wave pairing is attributed to the inter-band pair scattering arising from the intra-orbital AF spin $\bm s$ fluctuations. 
We also find that the Hund's coupling suppresses (enhances) the $d_{x^{2}-y^{2}}$-wave ($s_{\pm}$-wave pairing) pseudospin singlet pairing. 
Although the possibility of a pseudospin triplet pairing is considered, 
we find that it cannot be the most stable solution of the linearized Eliashberg equation with the maximum eigenvalue. 
We also argue that 
the enhancement of the $d_{x^{2}-y^{2}}$ pseudospin $j_{\text{eff}}=1/2$ singlet pairing with increasing the SOC $\lambda$ 
would be related to the purification of the orbital discussed in the context of cuprate superconductors~\cite{Sakakibara10, Sakakibara12, Sakakibara14}. 
We expect that the superconductivity with $s_{\pm}$-wave pairing symmetry found here is more relevant to hole doped Rh oxide Sr$_2$RhO$_4$. 

The rest of this paper is organized as follows. 
In Sec.~\ref{sec:model}, 
we introduce a $t_{2g}$ three-orbital Hubbard model with the SOC as an effective model for Sr$_2$IrO$_4$, 
and derive the RPA scheme and the linearized Eliashberg equation for this multi-band model with the SOC. 
We also introduce the pseudospin picture which is a better description in the large SOC limit. 
In Sec.~\ref{sec:results}, 
we show the numerical results to examine the possible superconductivity, 
including the pairing symmetry, and the mechanism mediated by different fluctuations. 
We also explore the parameter dependence of the superconductivity on the SOC and Hund's coupling. 
In Sec.~\ref{sec:summary}, we conclude the paper with summary and discussion. 

\section{Model and Formulation}\label{sec:model} 

\subsection{Effective model for Sr$_2$IrO$_4$} 

We consider an effective $t_{2g}$ three-orbital Hubbard model on the square lattice described by the following Hamiltonian: 
\begin{equation} 
H = H_{\text{kin}} +H_{\text{SO}} +H_{\text{int}} . 
\end{equation} 
The first term $H_{\text{kin}}$ is the kinetic energy term, 
\begin{equation}
H_{\text{kin}} = \sum_{\bm{k}} \sum_a \sum_\sigma (\epsilon^{a}_{\bm{k}} -\mu) c^{a\dagger}_{\bm{k}\sigma} c^{a}_{\bm{k}\sigma} , 
\end{equation} 
where 
\begin{equation} 
c^{a\dagger}_{\bm{k}\sigma} = \frac{1}{\sqrt{N}} \sum_{i} \mathrm{e}^{ \mathrm{i} \bm{k} \cdot \bm{R}_{i}} c^{a\dagger}_{i\sigma} 
\end{equation} 
and 
\begin{equation} 
c^{a}_{\bm{k}\sigma} = \frac{1}{\sqrt{N}} \sum_{i} \mathrm{e}^{-\mathrm{i} \bm{k} \cdot \bm{R}_{i}} c^{a}_{i\sigma}  
\end{equation} 
are the Fourier transformation of the creation and annihilation operators $c^{a\dagger}_{i\sigma}$ and $c^{a}_{i\sigma}$, 
respectively, of an electron in $t_{2g}$ orbital $a$ ($=yz, zx, xy$) with spin $\sigma$ ($=\uparrow , \downarrow$) 
at position $\bm{R}_{i}$ of site $i$ on the square lattice composed of $N$ sites. 
Here $\epsilon^{a}_{\bm{k}}$ represents the energy dispersion for $t_{2g}$ orbital $a$  and $\mu$ is the chemical potential. 
In this paper, we consider the simplest case in the absence of structural distortion, 
i.e., rotation of IrO$_6$ octahedra,~\cite{Randall57,Huang94} as in Ba$_2$IrO$_4$.~\cite{Okabe11,Boseggia13} 

The second term $H_{\text{SO}}$ is the SOC term with the coupling constant $\lambda$ described as 
\begin{equation} 
\begin{split} 
H_{\text{SO}} 
&=  \frac{\lambda}{2} \sum_{i} \sum_{a,b} \sum_{\sigma , \sigma^{\prime}} 
    \langle a | \bm{l} | b \rangle \cdot \langle \sigma | \bm{s} | \sigma^{\prime} \rangle c^{a\dagger}_{i\sigma} c^{b}_{i\sigma^{\prime}}  \\ 
&=  \frac{\lambda}{2} \sum_{i}\sum_\sigma 
    \left( \begin{array}{ccc} {c_{i\sigma}^{yz}}^{\dagger} \ \, {c_{i \sigma}^{zx}}^{\dagger} \ \, {c_{i\bar\sigma}^{xy}}^{\dagger} \end{array} \right)  \\
&\quad \times 
    \left( \begin{array}{ccc}  0           &  \mathrm{i} s_{\sigma} & -s_{\sigma}  \\
                              -\mathrm{i} s_{\sigma} &  0           &  \mathrm{i}           \\
                              - s_{\sigma} & -\mathrm{i}           &  0           \end{array} \right) 
    \left( \begin{array}{ccc}  c_{i\sigma}^{yz}  \\
                               c_{i\sigma}^{zx}  \\
                               c_{i\bar\sigma}^{xy} \end{array} \right) , 
\label{eq:hso} 
\end{split} 
\end{equation} 
where $\bm{l}$ ($\bm{s}$) is the orbital (spin) angular momentum operator, 
$\bar{\sigma}$ the opposite spin of $\sigma$, and $s_{\sigma}=1\,(-1)$ for $\sigma= \uparrow$ ($\downarrow$). 

The third term $H_{\text{int}}$ is the Coulomb interaction term, 
\begin{equation} 
\begin{split} 
H_{\text{int}} 
&=  U\sum_{i} \sum_{a} n^{a}_{i\uparrow}n^{a}_{i\downarrow} 
   +\frac{U^{\prime}}{2} \sum_{i} \sum_{a \ne b} n^{a}_{i} n^{b}_{i}  \\ 
&\quad 
   -J_{\text{H}} \sum_{i} \sum_{a \ne b} \left( {\bm S}^{a}_{i} \cdot{\bm S}^{b}_{i} +\frac{1}{4} n^{a}_{i} n^{b}_{i} \right)  \\ 
&\quad 
   +\frac{J^{\prime}}{2} \sum_{i} \sum_{a \ne b} c^{a\dagger}_{i\uparrow} c^{a\dagger}_{i\downarrow} c^{b}_{i\downarrow} c^{b}_{i\uparrow}, 
\end{split} 
\end{equation} 
which contains the intra- and inter-orbital on-site Hubbard interactions ($U$ and $U^{\prime}$, respectively), 
Hund's coupling ($J_{\rm H}$), and pair hopping ($J'$). 
Here $n^{a}_{i\sigma}=c^{a\dagger}_{i\sigma}c^{a}_{i\sigma}$, $n^{a}_i = n^{a}_{i\uparrow} + n^{a}_{i\downarrow}$, 
and  
$\bm{S}^{a}_{i} = \sum_{\sigma , \sigma^{\prime}} \langle \sigma | \bm{s} | \sigma^{\prime} \rangle c^{a\dagger}_{i\sigma} c^{b}_{i\sigma^{\prime}}$ is the spin operator at site $i$.  
We impose the constraint for $d$-orbital electrons in the atomic limit by setting 
$U=U^{\prime}+2J$ and $J=J_{\rm H}=J'$.~\cite{Kanamori63,Dagotto01} 
Although the realistic $U$ value for Ir oxides is estimated as large as 2--3 eV,~\cite{Arita12} 
we set smaller $U$ because the RPA calculation usually overestimates the instability toward ordered states. 

\subsection{Noninteracting limit}\label{sec:nonint} 

The one-body part of the Hamiltonian, $H_{0}=H_{\text{kin}} +H_{\text{SO}}$, is written in the matrix form as 
\begin{equation} 
\begin{split} 
H_{0} 
&=  \sum_{\bm{k}, \sigma} 
      \left( \begin{array}{ccc} c^{yz\dagger}_{\bm{k}\sigma} \ \, c^{zx\dagger}_{\bm{k}\sigma} \ \, c^{xy\dagger}_{\bm{k}\bar{\sigma}} \end{array} \right)  \\ 
&\quad \times 
      \left( \begin{array}{ccc}  \epsilon^{yz}_{\bm{k}}-\mu  &  \mathrm{i} s_{\sigma} \lambda/2       & -s_{\sigma}\lambda/2  \\ 
                                           -\mathrm{i} s_{\sigma}\lambda/2        &  \epsilon^{zx}_{\bm{k}}-\mu  &  \mathrm{i} \lambda/2           \\ 
                                           -s_{\sigma}\lambda/2        & -\mathrm{i} \lambda/2                  &  \epsilon^{xy}_{\bm{k}}-\mu \end{array} \right) 
      \left( \begin{array}{ccc} c^{yz}_{\bm{k}\sigma} \\ 
                                           c^{zx}_{\bm{k}\sigma} \\ 
                                           c^{xy}_{\bm{k}\bar{\sigma}} \end{array} \right) .  \label{eq:xi}
\end{split} 
\end{equation} 
According to the previous studies~\cite{Watanabe10,Watanabe13}, 
the energy band structure of Sr$_2$IrO$_4$ obtained by the first-principles calculation~\cite{Kim08} can be well reproduced 
using the following tight-binding energy dispersion for the three $t_{2g}$ orbitals: 
\begin{equation} 
\begin{split} 
\epsilon^{yz}_{\bm{k}} &= -2t_{5} \cos k_{x} -2t_{4} \cos k_{y},  \\ 
\epsilon^{zx}_{\bm{k}} &= -2t_{4} \cos k_{x} -2t_{5} \cos k_{y}, 
\end{split} 
\end{equation} 
and 
\begin{equation} 
\begin{split} 
\epsilon^{xy}_{\bm{k}} 
&= -2t_{1}( \cos k_{x} +\cos k_{y} ) -4t_{2}\cos k_{x} \cos k_{y}  \\ 
&\quad 
   -2t_{3}(\cos 2k_{x} +\cos 2k_{y}) +\mu_{xy} 
\end{split} 
\end{equation} 
with the parameters 
\begin{equation} 
\begin{split} 
&  ( t_{1}, t_{2}, t_{3}, t_{4}, t_{5}, \mu_{xy}, \lambda )  \\ 
&= (0.36, 0.18, 0.09, 0.37, 0.06, -0.36, 0.37) \, {\rm eV}.
\label{eq:parameter}
\end{split} 
\end{equation} 
Diagonalizing the ($3 \times 3$) Hamiltonian matrix in Eq.~(\ref{eq:xi}), 
we can obtain the three doubly degenerate energy band dispersions $E^{m}_{\bm{k}}$ ($m=1,2,3$). 
Note that each band is doubly degenerate because of the time-reversal symmetry and the inversion symmetry. 

\begin{figure*} 
\begin{center} 
\includegraphics[width=16.0cm,clip]{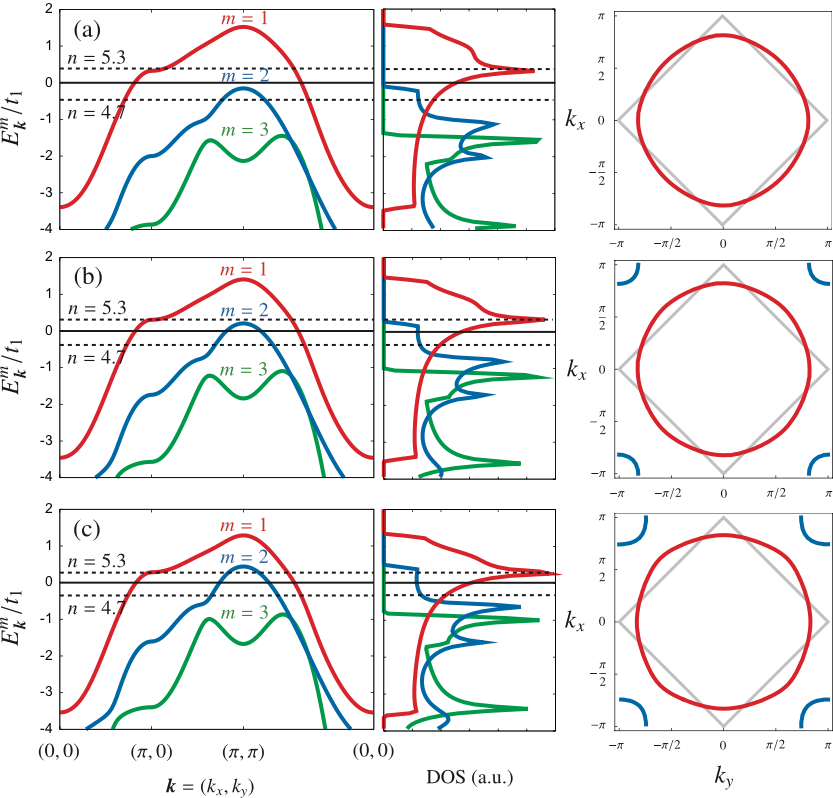} 
\caption{(color online) 
Energy band dispersion $E^{m}_{\bm k}$, DOS, and FS(s) of $H_0$
with (a) $\lambda=0.5 \ \text{eV} \sim 1.39 t_{1}$, (b) $\lambda= 0.37 \ \text{eV} \sim 1.03 t_{1}$, 
and (c) $\lambda=0.27 \ \text{eV} = 0.75 t_{1}$. 
The other parameters are given in Eq.~(\ref{eq:parameter}). 
In the left and middle panels, the solid horizontal lines represent the Fermi level for electron concentration $n=5$, 
and the dashed horizontal lines for $n=5.3$ and $n=4.7$. 
The red, blue, and green lines correspond to energy band $m=1$, 2, and 3, respectively. 
The FS(s) in the right panels is for $n=5$.  
The diamond-shaped squares (grey lines) in the right panels are guides for the eye. 
}
\label{fig:En_DOS_FS} 
\end{center} 
\end{figure*}

In this study, 
we consider the SOC $\lambda$, electron concentration $n$, and Hund's coupling $J$ as parameters 
to systematically study the possibility of superconductivity in Sr$_2$IrO$_4$. 
Figure~\ref{fig:En_DOS_FS} shows the energy band dispersion, the density of states (DOS), and Fermi surfaces (FSs) 
for $H_0$ with a parameter set given in Eq.~(\ref{eq:parameter}) but varying $\lambda$. 
The Fermi level $E_{\text{F}}$ for different $n$ is also indicated in Fig.~\ref{fig:En_DOS_FS}. 
The energy band basis that diagonalizes the Hamiltonian matrix in Eq.~(\ref{eq:xi}) 
is obtained with the $\bm{k}$-dependent ($6 \times 6$) unitary matrix $\mathcal{U}^{\dagger}_{\bm{k}}$ as 
\begin{equation} 
\begin{split} 
&  \left( \begin{array}{ccc} 
          d^{1\dagger}_{\bm{k}\Uparrow}    \ \,  d^{2\dagger}_{\bm{k}\Uparrow}    \ \,  d^{3\dagger}_{\bm{k}\Uparrow}  \ \, \big| \ \, 
          d^{1\dagger}_{\bm{k}\Downarrow}  \ \,  d^{2\dagger}_{\bm{k}\Downarrow}  \ \,  d^{3\dagger}_{\bm{k}\Downarrow}  \end{array} \right)  \\ 
&= \left( \begin{array}{ccc} 
          c^{yz\dagger}_{\bm{k}\uparrow}    \ \,  c^{zx\dagger}_{\bm{k}\uparrow}    \ \,  c^{xy\dagger}_{\bm{k}\uparrow}    \ \, \big| \ \, 
          c^{yz\dagger}_{\bm{k}\downarrow}  \ \,  c^{zx\dagger}_{\bm{k}\downarrow}  \ \,  c^{xy\dagger}_{\bm{k}\downarrow}  \end{array} \right) 
   \mathcal{U}^{\dagger}_{\bm{k}} , 
\end{split} 
\label{eq:band}
\end{equation} 
where $d^{m\dagger}_{\bm{k} \eta}$ and $d^{m}_{\bm{k} \eta}$ 
denote the creation and annihilation operator of an electron in the $m$-th band with 
$\eta\,(=\Uparrow, \Downarrow)$ indexing the doubly degenerate states. 

It is instructive to consider the basis set in the atomic limit, i.e., $\epsilon^{a}\to 0$ or $\lambda \rightarrow \infty$, which is achieved by diagonalizing $H_{\rm SO}$. 
This basis set is also the eigenstates of the pseudospin $\bm{j}_{\text{eff}} = -\bm{l} +\bm{s}$ and its $z$ component  with the eigenvalues $j_{\text{eff}}$ and $j^{z}_{\text{eff}}$, respectively, 
thus often called the effective total angular momentum $j_{\text{eff}}$ basis. 
This basis set is obtained by the site independent unitary transformation, 
\begin{equation} 
\begin{split} 
&  \left( \begin{array}{ccc}  
          a^{1\dagger}_{i\Uparrow}    \ \,  a^{2\dagger}_{i\Uparrow}    \ \,  a^{3\dagger}_{i\Uparrow}  \ \, \big| \ \, 
          a^{1\dagger}_{i\Downarrow}  \ \,  a^{2\dagger}_{i\Downarrow}  \ \,  a^{3\dagger}_{i\Downarrow}  \end{array} \right)  \\ 
&= \left( \begin{array}{ccc} 
          c^{yz\dagger}_{i\uparrow}    \ \,  c^{zx\dagger}_{i\uparrow}    \ \,  c^{xy\dagger}_{i\uparrow}  \ \, \big| \ \, 
          c^{yz\dagger}_{i\downarrow}  \ \,  c^{zx\dagger}_{i\downarrow}  \ \,  c^{xy\dagger}_{i\downarrow}  \end{array} \right) 
   \mathcal{W}^{\dagger}, 
\end{split} 
\label{eq:jeff}
\end{equation} 
where the $6 \times 6$ unitary matrix $\mathcal{W}^{\dagger}$ is given as 
\begin{equation} 
\mathcal{W}^{\dagger}= 
\left(  \begin{array}{ccc|ccc} 
        0 &  0 &  0 &  \frac{1}{\sqrt{3}} &  \frac{1}{\sqrt{2}} & -\frac{1}{\sqrt{6}}   \\ 
        0 &  0 &  0 & -\frac{\mathrm{i}}{\sqrt{3}} &  \frac{\mathrm{i}}{\sqrt{2}} &  \frac{i}{\sqrt{6}}  \\ 
        \frac{1}{\sqrt{3}} &  0 & -\frac{2}{\sqrt{6}} &  0 &  0 &  0  \\ \hline 
        \frac{1}{\sqrt{3}} & -\frac{1}{\sqrt{2}} &  \frac{1}{\sqrt{6}} &  0 &  0 &  0  \\ 
        \frac{\mathrm{i}}{\sqrt{3}} &  \frac{\mathrm{i}}{\sqrt{2}} &  \frac{\mathrm{i}}{\sqrt{6}} &  0 &  0 &  0  \\ 
        0 & 0 & 0 & -\frac{1}{\sqrt{3}} &  0 & -\frac{2}{\sqrt{6}}  \end{array} \right) . 
        \label{eq:uw}
\end{equation} 
The creation and annihilation operators $a^{m\dagger}_{i \eta}$ and $a^{m}_{i \eta}$ in the $j_{\text{eff}}$ basis 
with $(m, \eta)= (1, \Uparrow \Downarrow)$, $(2, \Uparrow \Downarrow)$, and $(3, \Uparrow \Downarrow)$ correspond to 
$(j_{\text{eff}}, j^{z}_{\text{eff}})=(1/2, \pm 1/2)$, $(3/2, \mp 3/2)$, and $(3/2, \pm 1/2)$, respectively. 
In the atomic limit, these two unitary matrices becomes the same, i.e.,  
$ \mathcal{U}^{\dagger}_{\bm{k}} \to \mathcal{W}^{\dagger}$, 
and the energy band eigenstates described by $d^{m\dagger}_{\bm{k} \eta}$ and $d^{m}_{\bm{k} \eta}$ with 
$(m,\eta) = (1, \Uparrow \Downarrow)$, $(2, \Uparrow \Downarrow)$, and $(3, \Uparrow \Downarrow)$ 
approach to the $j_{\text{eff}}$ bases with 
$(j_{\rm eff},j_{\rm eff}^z)=(1/2,\pm1/2)$, $(3/2,\mp3/2)$, and $(3/2,\pm1/2)$. 

\begin{figure*} 
\begin{center} 
\includegraphics[width=17.0cm,clip]{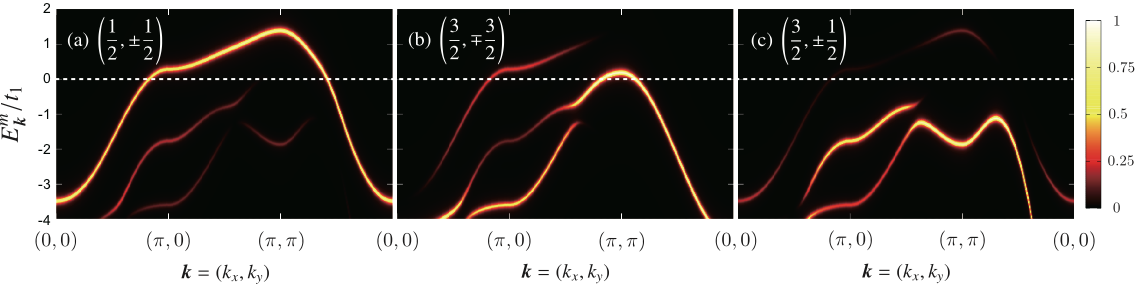} 
\caption{(color online) 
Intensity plots of the projected single-particle excitation spectra $A^{(j_{\rm eff},j_{\rm eff}^z)}(\bm{k},\omega)$ 
onto the $j_{\text{eff}}$ bases in the noninteracting limit with $\lambda= 0.37 \ \text{eV} \sim 1.03 t_{1}$: 
(a) $(j_{\rm eff},j_{\rm eff}^z)=(1/2,\pm1/2)$, (b) $(3/2,\mp3/2)$, and (c) $(3/2,\pm1/2)$. 
The intensity is normalized by the maximum value $A_{\text{max}} = \max \big[ A^{(j_{\rm eff},j_{\rm eff}^z)}(\bm{k},\omega) \big]$ for each figure. 
The Fermi level at $n=5$ is indicated by white dashed lines. 
The model parameters are given in Eq.~(\ref{eq:parameter}). 
} 
\label{fig:jeff} 
\end{center} 
\end{figure*}

In fact, we can observe in Fig.~\ref{fig:jeff} that the energy band $E^{m}_{\bm{k}}$ with $m=1$, 
which crosses the Fermi level at $n=5$, forming electron-like FS, and is the most important for the later discussion, 
is approximately described by the $j_{\text{eff}}=1/2$ basis when $\lambda= 0.37 \ \text{eV} \sim 1.03 t_{1}$, relevant to Sr$_2$IrO$_4$, 
because almost all the spectral weight projected onto the $j_{\text{eff}}=1/2$ basis (i.e., partial spectral function) is 
concentrated on the band $E^{m}_{\bm{k}}$ with $m=1$. 
On the other hand, the spectral weight projected onto the $j_{\text{eff}}=3/2$ bases 
are concentrated mostly on the bands $E^{m}_{\bm{k}}$ with $m=2$ and 3. 
We can also notice that 
the band $E^{m}_{\bm{k}}$ with $m=2$ crosses the Fermi level at $n=5$ with this $\lambda$ value, 
forming the hole-like FS around $\bm{k}=(\pi,\pi)$ [see Fig.~\ref{fig:En_DOS_FS}(b)], which has the strong character of $(j_{\text{eff}}, j^{z}_{\text{eff}})=(3/2, \pm 3/2)$. 

\begin{figure} 
\begin{center} 
\includegraphics[width=8.5cm,clip]{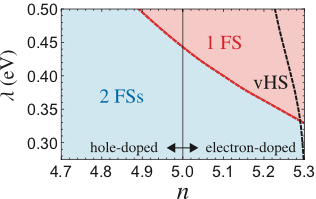} 
\caption{(color online) 
The number of FSs crossing the Fermi level with varying the SOC $\lambda$ and the electron concentration $n$ in the noninteracting limit. 
There exists one (two) FS(s) in the region indicated by red (blue). 
The black lines represents $(\lambda,n)$ where the van Hove singularity (vHS) is located exactly at the Fermi level. 
The hopping parameters are set as in Eq.~(\ref{eq:parameter}). 
} 
\label{fig:lam_n}
\end{center} 
\end{figure}

The number of FSs crossing the Fermi level depends on the SOC $\lambda$ as well as the electron concentration $n$. 
Since the SOC tends to increase the separation between the $j_{\rm eff}=1/2$ based band and the $j_{\rm eff}=3/2$ 
based bands, there is only one FS crossing the Fermi level for the larger SOC when $n$ is around 5, 
as shown in Fig.~\ref{fig:En_DOS_FS}. 
It is also apparent that when the number of electron increases, 
only the highest band, i.e., the $j_{\rm eff}=1/2$ based band, is relevant and thus there is only a single FS crossing the Fermi level (see Fig.~\ref{fig:En_DOS_FS}). 
This feature is summarized in Fig.~\ref{fig:lam_n} for the case of the tight-binding parameters given in Eq.~(\ref{eq:parameter}) 
except that here the SOC $\lambda$ is varied. 
In Fig.~\ref{fig:lam_n}, 
we also indicate $(\lambda,n)$ at which the van Hove singularity (vHS) is located exactly at the Fermi level. 
The vHS is originated from the inflection point of the energy band $E^{m}_{\bm{k}}$ with $m=1$ at $\bm{k}= (\pm\pi , 0)$ and $(0,\pm\pi)$, 
which is always located in the electron-doped side for the parameters considered here. 
  

\subsection{Linearized Eliashberg equation in RPA} 

Here, we show the procedure of the RPA calculations to treat the electron correlations. 
The RPA incorporates the spin and charge fluctuations by including the infinite summation of bubble and ladder diagrams~\cite{BSW89, Bickers89, Dahm95, Takimoto04, Yada05}.  
Since our system $H$ is a multi-band system with the SOC, 
it requires a spin-dependent multi-band formalism,~\cite{KN14} 
in which the spin SU(2) symmetry is broken due to the presence of the SOC. 

Let us first derive the linearized Eliashberg equation for a SC gap function. 
The SC properties such as the pairing symmetry and the transition temperature 
can be determined by solving the linearized Eliashberg equation. 
For this purpose, 
we first define the normal Green's functions $\hat{G}(k)$ and $\hat{\bar{G}}(k)$ in the multi-band system as 
\begin{equation} 
\begin{split} 
G_{\alpha \beta}(k) 
&= -\int^{\beta}_{0}d\tau \ \mathrm{e}^{\mathrm{i} \omega_{n}\tau} \bra T_{\tau} c_{\bm{k}\alpha}(\tau) c^{\dagger}_{\bm{k}\beta}(0) \ket ,  \\ 
\bar{G}_{\alpha \beta}(k) 
&= -\int^{\beta}_{0}d\tau \ \mathrm{e}^{\mathrm{i} \omega_{n}\tau} \bra T_{\tau} c^{\dagger}_{-\bm{k}\alpha}(\tau) c_{-\bm{k}\beta}(0) \ket ,
\end{split} 
\end{equation} 
where $k=(\bm{k}, i\omega_{n})$ denotes a set of wave vector $\bm{k}$ and Matsubara frequency 
$\omega_{n} = (2n+1)\pi /\beta$ for fermions ($n$: integer), 
and $T$ ($\beta = 1/k_{\text{B}}T$) denotes the (inverse) temperature. 
Note that we temporarily treat the Greek indices as a set of the orbital and spin degrees of freedom: 
$\alpha = (a, \sigma)$ contains the orbital index $a = yz,zx, xy$ and the spin index $\sigma = \uparrow, \downarrow$, 
and thus $c_{\bm{k}\alpha} = c^{a}_{\bm{k}\sigma}$ and $c^{\dagger}_{\bm{k}\alpha} = c^{a\dagger}_{\bm{k}\sigma}$. 
The time dependence of the field operators in the imaginary-time $\tau$ formalism is defined as 
$c_{\bm{k}\alpha}(\tau) = \mathrm{e}^{H\tau} c_{\bm{k}\alpha} \mathrm{e}^{-H\tau}$ 
and $c^{\dagger}_{\bm{k}\alpha}(\tau) = \mathrm{e}^{H\tau} c^{\dagger}_{\bm{k}\alpha} \mathrm{e}^{-H\tau}$. 
$\bra \cdots \ket = \mathrm{Tr} [\mathrm{e}^{-\beta H} \cdots] /Z$ is the statistical average of the system $H$ 
and $Z= \mathrm{Tr} \ \mathrm{e}^{-\beta H}$ is the partition function. 
From the definition of $\hat{G}(k)$ and $\hat{\bar{G}}(k)$, we can easily show that 
$\bar{G}_{\alpha \beta}(k) = -G_{\beta \alpha}(-k)$.  

The noninteracting Green's function is obtained from the one-body dispersion matrix $\hat{\xi}_{\bm{k}}$ as 
\begin{equation} 
\hat{G}_{0}(k)= \Big[ \mathrm{i} \omega_{n}-\hat{\xi}_{\bm{k}} \Big]^{-1} , 
\end{equation} 
where the matrix elements of $\hat{\xi}_{\bm{k}}$ can be given by rewriting the one-body Hamiltonian $H_{0}$ 
in this representation as 
\begin{equation} 
\begin{split} 
H_{0} 
&=  \sum_{\bm{k}} \sum_{\alpha , \beta} c^{\dagger}_{\bm{k} \alpha} \xi_{\bm{k} \, \alpha \beta} c_{\bm{k} \beta}  \\ 
&=  \sum_{\bm{k}} 
    \left(  \begin{array}{ccc} 
            c^{yz\dagger}_{\bm{k}\uparrow}    \ \,  c^{zx\dagger}_{\bm{k}\uparrow}    \ \,  c^{xy\dagger}_{\bm{k}\uparrow}  \ \, \big| \ \, 
            c^{yz\dagger}_{\bm{k}\downarrow}  \ \,  c^{zx\dagger}_{\bm{k}\downarrow}  \ \,  c^{xy\dagger}_{\bm{k}\downarrow}  \end{array} \right)  \\ 
&\times 
    \left( \begin{array}{ccc|ccc} 
    \xi^{yz}_{\bm k} & \mathrm{i} \frac{\lambda}{2} & 0 & 0 & 0 & -\frac{\lambda}{2} \\
   -\mathrm{i} \frac{\lambda}{2} & \xi^{zx}_{\bm k} & 0 & 0 & 0 & \mathrm{i} \frac{\lambda}{2} \\
    0 & 0 & \xi_{\bm{k}}^{xy} & \frac{\lambda}{2} & -\mathrm{i} \frac{\lambda}{2} & 0 \\ \hline 
    0 & 0 & \frac{\lambda}{2} & \xi_{\bm{k}}^{yz} & -\mathrm{i} \frac{\lambda}{2} & 0 \\
    0 & 0 & \mathrm{i}\frac{\lambda}{2} & {\rm i}\frac{\lambda}{2} & \xi_{\bm{k}}^{zx} & 0 \\
   -\frac{\lambda}{2} & -\mathrm{i} \frac{\lambda}{2} & 0 & 0 & 0 & \xi_{\bm k}^{xy}  \end{array} \right) 
    \left(  \begin{array}{c} 
            c^{yz}_{\bm{k}\uparrow}    \\  c^{zx}_{\bm{k}\uparrow}    \\  c^{xy}_{\bm{k}\uparrow}  \\ \hline 
            c^{yz}_{\bm{k}\downarrow}  \\  c^{zx}_{\bm{k}\downarrow}  \\  c^{xy}_{\bm{k}\downarrow}  \end{array} \right).  
\end{split} 
\end{equation} 
The noninteracting Green's functions also satisfy that $\bar{G}_{0 \, \alpha \beta}(k) = -G_{0 \, \beta \alpha}(-k)$. 

We also define the anomalous Green's functions $\hat{F}(k)$ and $\hat{\bar{F}}(k)$ in the multi-band system as 
\begin{equation} 
\begin{split} 
F_{\alpha \beta}(k) 
&= -\int^{\beta}_{0}d\tau \ \mathrm{e}^{\mathrm{i} \omega_{n}\tau} \bra T_{\tau} c_{\bm{k}\alpha}(\tau) c_{-\bm{k}\beta}(0) \ket ,  \\ 
\bar{F}_{\alpha \beta}(k) 
&= -\int^{\beta}_{0}d\tau \ \mathrm{e}^{\mathrm{i} \omega_{n}\tau} \bra T_{\tau} c^{\dagger}_{-\bm{k}\alpha}(\tau) c^{\dagger}_{\bm{k}\beta}(0) \ket . 
\end{split} 
\end{equation} 
The SC order parameter, i.e., 
the SC gap function $\hat{\Delta}(k)$ can be interpreted as the self-energy of the anomalous Green's function 
$\hat{F}(k)$. 
From the Dyson's equation for $\hat{G}(k)$ and $\hat{F}(k)$, 
we can obtain the following relation: 
\begin{equation} 
F_{\alpha \beta}(k) 
=  \sum_{\alpha^{\prime}, \beta^{\prime}} G_{\alpha \alpha^{\prime}}(k) \Delta_{\alpha^{\prime} \beta^{\prime}}(k) \bar{G}_{\beta^{\prime}\beta}(k) , 
\label{eq:FDyson}
\end{equation} 
where we assume that the temperature is around the critical temperature $T_{\text{c}}$ 
and thus $\hat{F}(k)$ and $\hat{\Delta}(k)$ are infinitesimally small.  

\begin{figure}[t]
\begin{center}
\includegraphics[width=6.0cm,clip]{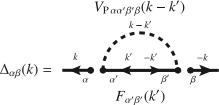} 
\caption{ 
Diagrammatic representation of the SC gap function $\hat{\Delta}(k)$. 
$\hat{\Delta}(k)$ corresponds to the self-energy of the anomalous Green's function $\hat{F}(k)$ 
with the SC pairing interaction $\hat{V}_{\text{P}}(q)$ to form the Cooper pairs. 
} 
\label{fig:gap}
\end{center} 
\end{figure}

The SC gap function $\hat{\Delta}(k)$ can be obtained from the functional derivative of the Luttinger--Ward functional $\Phi$ 
with respect to the anomalous Green's function $\hat{\bar{F}}(k)$~\cite{Baym61, Baym62,Yanase03}. 
Here the SC gap function can be given as 
\begin{equation} 
\Delta_{\alpha \beta}(k) 
=  \frac{1}{N\beta}\sum_{k^{\prime}} \sum_{\alpha^{\prime}\beta^{\prime}} 
    V_{\text{P} \, \alpha \alpha^{\prime} \beta^{\prime} \beta}(k-k^{\prime}) 
    F_{\alpha^{\prime} \beta^{\prime}}(k^{\prime}) , 
\label{eq:Gapdef} 
\end{equation} 
where 
$\hat{V}_{\text{P}}(q)$ represents the SC pairing interaction, 
i.e.,  the effective interaction that induces the self-energy of the anomalous Green's function 
$\hat{F}(k)$ (see Fig. \ref{fig:gap}), and $q=(\bm{q}, i\varepsilon_{m})$ with 
$\varepsilon_{m} = 2m \pi /\beta$ being a Matsubara frequency for bosons ($m$: integer). 
$\hat{V}_{\text{P}}(q)$ can be uniquely determined once the Luttinger--Ward functional $\Phi$ is chosen. 
In the RPA, 
the infinite summation of bubble and ladder diagrams connected to each other 
via the Coulomb interactions are taken into account in $\Phi$. 

By combining Eqs.~(\ref{eq:FDyson}) and (\ref{eq:Gapdef}), 
we can obtain the linearized Eliashberg equation. 
To solve this equation, 
first we neglect the frequency dependence of the gap function, i.e., $\hat{\Delta}(k) \rightarrow \hat{\Delta}(\bm{k})$, 
and the pairing interaction, i.e., $\hat{V}_{\text{P}}(q) \rightarrow \hat{V}_{\text{P}}(\bm{q}) \equiv \hat{V}_{\text{P}}(\bm{q}, i\varepsilon_{m}=0)$, for simplicity. 
In addition to this, in the RPA calculation, 
the Green's functions $\hat{G}(k)$ and $\hat{\bar{G}}(k)$ are replaced 
with the noninteracting Green's functions $\hat{G}_{0}(k)$ and $\hat{\bar{G}}_{0}(k)$, respectively. 
Thus, the linearized Eliashberg equation can be given as  
\begin{equation} 
\begin{split} 
\lambda_{\text{e}} \Delta_{\alpha \beta}(\bm{k}) 
&= -\frac{1}{N} \sum_{\bm{k}^{\prime}} \sum_{\alpha^{\prime}\beta^{\prime}} \sum_{\mu \nu} 
       V_{\text{P} \, \alpha \alpha^{\prime}\beta^{\prime} \beta}(\bm{k}-\bm{k}^{\prime})  \\ 
&\quad \times 
       \phi_{\alpha^{\prime} \beta^{\prime} \mu \nu}(\bm{k}^{\prime}) 
       \Delta_{\mu \nu}(\bm{k}^{\prime}), 
\label{eq:GapEq} 
\end{split} 
\end{equation} 
where the summation over the Matsubara frequencies $\omega_{n}$ is gathered 
in $\hat{\phi}$ 
as 
\begin{equation}
\phi_{\alpha^{\prime} \beta^{\prime} \mu \nu}(\bm{k}^{\prime}) 
= -\frac{1}{\beta} \sum_{\mathrm{i} \omega_{n}} 
    G_{0 \, \alpha^{\prime} \mu}( \bm{k}^{\prime}, \mathrm{i} \omega_{n}) 
    \bar{G}_{0 \, \nu \beta^{\prime}}(\bm{k}^{\prime}, \mathrm{i} \omega_{n}) . 
\end{equation} 
Note that we have to impose the constraint  
\begin{equation}
\Delta^{ab}_{\alpha \beta}(\bm{k})=-\Delta^{ba}_{\beta \alpha}(-\bm{k})  
\end{equation} 
on the solution $\hat{\Delta}(\bm{k})$ of the linearized Eliashberg equation in Eq.~(\ref{eq:GapEq})
because of the anticommutation relation for fermions. 
Note also that the mixture of singlet and triplet pairing is permitted as a consequence of the spin SU(2) symmetry breaking due to a finite SOC. 
Moreover, it is convenient to consider the eigenvalue $\lambda_{\text{e}}$ of the linearized Eliashberg equation in Eq.~(\ref{eq:GapEq}). 
The SC transition occurs when $\lambda_{\text{e}}=1$ at the SC critical temperature $T_{\text{c}}$. 
The value of $\lambda_{\text{e}}$ itself gives us a criterion for the strength of the superconductivity at temperature 
$T$ around $T_{\text{c}}$. 

To obtain the SC paring interaction $\hat{V}_{\text{P}}(q)$, 
we next consider the dynamical correlation function $\hat{\chi}(q)$ defined as 
\begin{equation} 
\begin{split} 
\chi_{\alpha \beta \gamma \delta}(q) 
&=  \frac{1}{N} \int^{\beta}_{0} d\tau \ \mathrm{e}^{\mathrm{i} \epsilon_{m} \tau}  \\ 
&\quad \times 
       \sum_{\bm{k}, \bm{k}^{\prime}} 
       \bra T_{\tau} c^{\dagger}_{\bm{k} \beta}(\tau) c_{\bm{k}+\bm{q} \alpha}(\tau) 
                            c^{\dagger}_{\bm{k}^{\prime} \gamma}(0) c_{\bm{k}^{\prime}-\bm{q} \delta}(0) \ket_{\text{c}} , 
\end{split} 
\end{equation} 
where the subscript ``c" indicates that only connected diagrams are included. 
In the weak-coupling theory, 
$\hat{\chi}(q)$ can be evaluated from the infinite summation of the bubble and ladder diagrams connected to each other via the Coulomb interactions, 
which is a solution of the Bethe--Salpeter equation for the particle-hole correlation function. 
Furthermore, in the RPA calculations,  
the bubble and ladder diagrams are replaced with those composed of the noninteracting Green's functions $\hat{G}_{0}(k)$. 
One single bubble/ladder diagram in the RPA thus corresponds to the noninteracting dynamical correlation function, 
\begin{equation}
\chi_{0 \, \alpha \beta \gamma \delta}(q) = -\frac{1}{N\beta} \sum_{k} G_{0 \, \alpha \gamma}(q+k) G_{0 \, \delta \beta}(k) , 
\end{equation} 
and this bubble/ladder diagram is connected to others repeatedly via the Coulomb interaction $\hat{V}$ in $\hat{\chi}(q)$. 

Showing explicitly the orbital and spin indices, 
we can now express the noninteracting dynamical correlation function 
$\chi_{0 \, \alpha \beta \gamma \delta}(q) = \chi^{abcd}_{0 \, \sigma_{1} \sigma_{2} \sigma_{3} \sigma_{4}}(q)$ 
in a $2^{2} \times 2^{2}$ block matrix form with respect to the spin indices (the number of spin indices being $2$), 
each element of the block matrix being composed of  $\hat{\chi}_{0 \, \sigma_{1} \sigma_{2} \sigma_{3} \sigma_{4}}(q)$, i.e.,  
\begin{equation} 
\hat{\chi}_{0}(q) = 
\begin{pmatrix} 
  \hat{\chi}_{0 \ \uparrow   \uparrow   \uparrow   \uparrow} & \hat{\chi}_{0 \ \uparrow   \uparrow   \uparrow   \downarrow} 
& \hat{\chi}_{0 \ \uparrow   \uparrow   \downarrow \uparrow} & \hat{\chi}_{0 \ \uparrow   \uparrow   \downarrow \downarrow}  \\ 
  \hat{\chi}_{0 \ \uparrow   \downarrow \uparrow   \uparrow} & \hat{\chi}_{0 \ \uparrow   \downarrow \uparrow   \downarrow} 
& \hat{\chi}_{0 \ \uparrow   \downarrow \downarrow \uparrow} & \hat{\chi}_{0 \ \uparrow   \downarrow \downarrow \downarrow}  \\ 
  \hat{\chi}_{0 \ \downarrow \uparrow   \uparrow   \uparrow} & \hat{\chi}_{0 \ \downarrow \uparrow   \uparrow   \downarrow} 
& \hat{\chi}_{0 \ \downarrow \uparrow   \downarrow \uparrow} & \hat{\chi}_{0 \ \downarrow \uparrow   \downarrow \downarrow}  \\ 
  \hat{\chi}_{0 \ \downarrow \downarrow \uparrow   \uparrow} & \hat{\chi}_{0 \ \downarrow \downarrow \uparrow   \downarrow} 
& \hat{\chi}_{0 \ \downarrow \downarrow \downarrow \uparrow} & \hat{\chi}_{0 \ \downarrow \downarrow \downarrow \downarrow} 
\end{pmatrix}(q) ,  
\end{equation} 
where the four rows (columns) correspond to 
$\sigma_{1} \sigma_{2} \,\, (\sigma_{3} \sigma_{4})=\uparrow \uparrow, \, \uparrow \downarrow, \,\downarrow \uparrow, \,\downarrow \downarrow$. 
Each element of the block matrix, $\hat{\chi}_{0 \, \sigma_{1} \sigma_{2} \sigma_{3} \sigma_{4}}(q)$, 
is a $3^{2} \times 3^{2}$ matrix $\chi^{abcd}_{0 \, \sigma_{1} \sigma_{2} \sigma_{3} \sigma_{4}}(q)$
with respect to the orbital indices (the number of orbital indices being $3$ in the three-orbital case), 
where the nine rows (columns) correspond to $ab \,\, (cd) = yz \ yz, \,yz \ zx, \,yz \ xy, \,zx \ yz, \,zx \ zx, \,zx \ xy, \,xy \ yz, \,xy \ zx, \,xy \ xy$. 

In the same way, the Coulomb interaction $\hat{V}$ can also be expressed 
in a $2^{2} \times 2^{2}$ block matrix form, composed of $\hat{V}_{\sigma_{1} \sigma_{2} \sigma_{3} \sigma_{4}}$, 
with respect to the spin indices, i.e.,   
\begin{equation} 
\hat{V} = 
\begin{pmatrix} 
\hat{V}_{\uparrow \uparrow \uparrow \uparrow}  &  0  &  0  &  \hat{V}_{\uparrow \uparrow \downarrow \downarrow}  \\ 
0  &  \hat{V}_{\uparrow \downarrow \uparrow \downarrow} &  0  &  0  \\ 
0  &  0  & \hat{V}_{\downarrow \uparrow \downarrow \uparrow} &  0  \\ 
\hat{V}_{\downarrow \downarrow \uparrow \uparrow}  &  0  &  0  &  \hat{V}_{\downarrow \downarrow \downarrow \downarrow} 
\end{pmatrix} , 
\end{equation} 
where each element of the block matrix, $\hat{V}_{\sigma_{1} \sigma_{2} \sigma_{3} \sigma_{4}}$, can be given explicitly as 
\begin{equation} 
\begin{split} 
&V^{abcd}_{\uparrow \uparrow \uparrow \uparrow}=V^{abcd}_{\downarrow \downarrow \downarrow \downarrow}= 
\begin{cases} 
U^{\prime}-J     & (a=c \neq b=d)  \\ 
-(U^{\prime}-J)  & (a=b \neq c=d)  \\ 
0                & (\text{otherwise})
\end{cases},  \\ 
&V^{abcd}_{\uparrow \uparrow \downarrow \downarrow}=V^{abcd}_{\downarrow \downarrow \uparrow \uparrow}= 
\begin{cases} 
-U           & (a=b=c=d)       \\ 
-J           & (a=c \neq b=d)  \\ 
-U^{\prime}  & (a=b \neq c=d)  \\ 
-J           & (a=d \neq b=c)  \\ 
0            & (\text{otherwise})
\end{cases},  \\ 
&V^{abcd}_{\uparrow \downarrow \uparrow \downarrow}=V^{abcd}_{\downarrow \uparrow \downarrow \uparrow}= 
\begin{cases} 
U           & (a=b=c=d)        \\ 
U^{\prime}  & (a=c \neq b=d)   \\ 
J           & (a=b \neq c=d)   \\ 
J           & (a=d \neq b=c)   \\ 
0           & (\text{otherwise})
\end{cases}, 
\end{split} 
\end{equation} 
and other elements of the block matrix are zero. 

Therefore, these matrices $\hat{\chi}_{0}(q)$ and $\hat{V}$ can be multiplied simply by the matrix (or more precisely tensor) product 
with respect to the spin and orbital indices, i.e., 
\begin{equation}
[\hat{A} \hat{B}]^{abcd}_{\sigma_{1} \sigma_{2} \sigma_{3} \sigma_{4}} 
=  \sum_{m, n} \sum_{\sigma , \sigma^{\prime}} 
    A^{abmn}_{\sigma_{1} \sigma_{2} \sigma \sigma^{\prime}} B^{mncd}_{\sigma \sigma^{\prime} \sigma_{3} \sigma_{4}} . 
\end{equation} 
Thus, the dynamical correlation function $\hat{\chi}(q)$ in the RPA, which is the infinite series 
$\hat{\chi}(q) = \hat{\chi}_{0}(q) +\hat{\chi}_{0}(q)\hat{V}\hat{\chi}_{0}(q) +\cdots = \hat{\chi}_{0}(q) +\hat{\chi}_{0}(q)\hat{V}\hat{\chi}(q)$, 
can be given as 
\begin{equation} 
\hat{\chi}(q) = \left( 1 -\hat{\chi}_{0}(q) \hat{V} \right)^{-1} \hat{\chi}_{0}(q) . 
\end{equation} 
In the RPA, the superconductivity is mediated by spin and orbital (charge) fluctuations that 
are included in the dynamical correlation function $\hat{\chi}(q)$. 
Therefore, the SC pairing interaction $\hat{V}_{\text{P}}(q)$ is finally given as  
\begin{equation} 
\hat{V}_{\text{P}}(q) 
= -\Bigg[ \frac{\hat{V}}{2} +\hat{V} \hat{\chi}(q) \hat{V} \Bigg] , 
\label{eq:VP} 
\end{equation} 
where the factor $1/2$ in the first term is required in order to avoid the double counting in the first order diagrams.

\section{Numerical Results} \label{sec:results} 

To investigate the possible superconductivity in the $t_{2g}$ three-orbital Hubbard model with the SOC, an effective model for Sr$_2$IrO$_4$, 
we solve the linearized Eliashberg equation numerically with the RPA. 
In the numerical calculations, 
we set the number of momentum mesh points in the first Brillouin zone as large as $64\times64$ points 
and the temperature $T=0.03$ eV $\sim 300$ K. 

\subsection{Possible Superconductivity} \label{sec:PS}

For the analysis of possible superconductivity, 
it is useful to represent the SC gap function $\hat{\Delta}(\bm{k})$, 
represented in terms of the $t_{2g}$ basis in Eq.~(\ref{eq:GapEq}), 
by using the energy band basis $m$ that diagonalizes the noninteracting Hamiltonian $H_0$, i.e., 
\begin{equation} 
\hat{D}(\bm{k}) =  \hat{\mathcal{U}}_{\bm{k}} \hat{\Delta}(\bm{k}) \hat{\mathcal{U}}^{\mathrm{T}}_{-\bm{k}}, 
\end{equation} 
where the unitary matrix $\hat{\mathcal{U}}_{\bm{k}}$ is introduced in Eq.~(\ref{eq:band}). 
Recall that the pseudospin ($= \Uparrow , \Downarrow $), specifying the doubly degenerated energy band, 
is a good quantum number, not the electron spin $\sigma\,(\uparrow,\downarrow)$, because of the presence of the SOC. 
Since there exists the spatial inversion symmetry in the Hamiltonian $H$, 
the SC gap function $\hat{D}(\bm{k})$ represented in terms of the energy band basis $m$ 
can be characterized as the ``pseudospin" singlet or triplet pairing. 
This is an analogy to the intrinsic spin singlet or triplet pairing when the SOC is absent. 
In the numerical calculations, 
we can distinguish the pseudospin singlet and triplet pairings by imposing the even or odd parity on the SC gap function, i.e.,  
\begin{equation} 
\hat{D}(-\bm{k}) = \pm \left[\hat{D}(\bm{k})\right]^{\rm T} . 
\label{eq:SCG}
\end{equation} 
Note that this is equivalent to $\hat{\Delta}(-\bm{k}) = \pm \left[\hat{\Delta}(\bm{k})\right]^T$. 

\begin{figure}
\begin{center}
\includegraphics[width=7.5cm,clip]{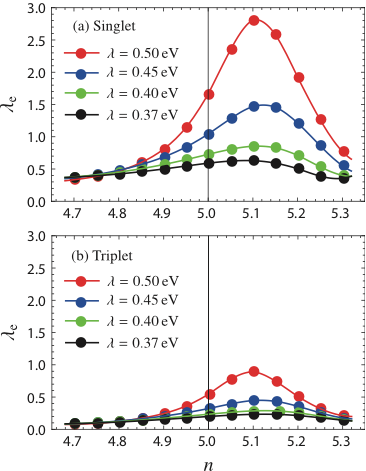} 
\caption{(color online) 
The largest eigenvalue $\lambda_{\text{e}}$ of the linearized Eliashberg equation with 
(a) the pseudospin singlet pairing and (b) the pseudospin triplet pairing vs. electron concentration $n$ for $U=0.75$ eV, $J/U=0.05$, 
and different values of $\lambda$ indicated. 
The other parameters are given in Eq.~(\ref{eq:parameter}). 
The most stable singlet (triplet) pairing is $d_{x^{2}-y^{2}}$-wave ($p$-wave) symmetry (also see Fig.~\ref{fig:gap1}). 
} 
\label{fig:EV_dt}
\end{center} 
\end{figure}

\begin{figure}
\begin{center}
\includegraphics[width=7.5cm,clip]{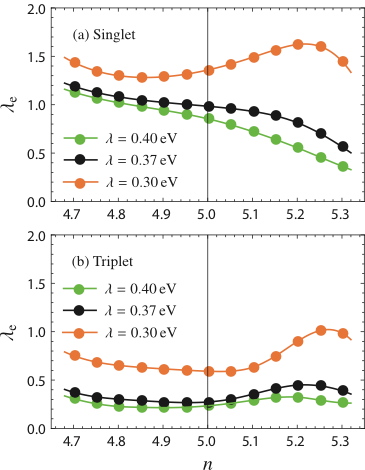} 
\caption{(color online) 
Same as Fig.~\ref{fig:EV_dt} but for $U=0.9$ eV and $J/U=0.3$. 
The most stable singlet (triplet) pairing is $s_{\pm}$-wave ($p_\pm$-wave) symmetry (also see Fig.~\ref{fig:gap2}). 
} 
\label{fig:EV_st}
\end{center} 
\end{figure}

Figure~\ref{fig:EV_dt} and Figure~\ref{fig:EV_st} show the results of the largest eigenvalue $\lambda_{\text{e}}$ of the linearized Eliashberg equation 
with the pseudospin singlet and triplet pairings for the representative two sets of the Coulomb interactions $U$ and $J$. 
We find that 
the pseudospin singlet pairing is always favored as compared to the pseudospin triplet pairing for all the parameter sets studied here 
because the maximum eigenvalue $\lambda_{\text{e}}$ for the pseudospin singlet pairing is always larger than that for the pseudospin triplet pairing. 
It is also interesting to notice that 
the $\lambda$ dependence of the eigenvalue $\lambda_{\text{e}}$ is opposite in Fig.~\ref{fig:EV_dt} and Fig.~\ref{fig:EV_st}. 
The eigenvalue $\lambda_{\text{e}}$ increases (decreases) with $\lambda$ for a given electron concentration $n$ in Fig.~\ref{fig:EV_dt} (Fig.~\ref{fig:EV_st}) 
where a relatively small (large) $J$ is set. 


\begin{figure}
\begin{center}
\includegraphics[width=8.0cm,clip]{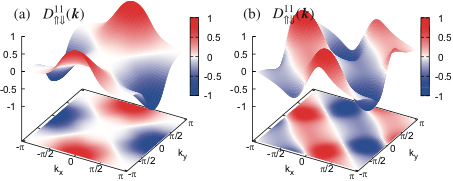} 
\caption{(color online) 
SC gap functions at the $m=1$ energy band for (a) the pseudospin singlet pairing and (b) the pseudospin triplet pairing, 
clearly showing $d_{x^{2}-y^{2}}$-wave symmetry and $p$-wave symmetry, respectively. 
We set $\lambda = 0.5$ eV and $n=5.1$, and the other parameters are the same as in Fig.~\ref{fig:EV_dt}. 
} 
\label{fig:gap1}
\end{center} 
\end{figure}

Furthermore, as shown in Fig.~\ref{fig:gap1}, we find that 
the dominant SC gap function has $d_{x^{2}-y^{2}}$-wave symmetry for the pseudospin singlet pairing and $p$-wave symmetry for the pseudospin triplet pairing 
in the parameter region shown in Fig.~\ref{fig:EV_dt}.  
These pairings are formed mainly in the $m=1$ energy band that has a strong character of $j_{\rm eff}=1/2$ (see Fig.~\ref{fig:jeff}). 
These pairings are much favored for larger $\lambda$ and particularly in the electron doped region. 
This is because the single band nature is enhanced with increasing $\lambda$ in the electron doped region 
and also because there is the vHS in the electron-doped side (see Fig.~\ref{fig:lam_n}).

\begin{figure}
\begin{center}
\includegraphics[width=8.0cm,clip]{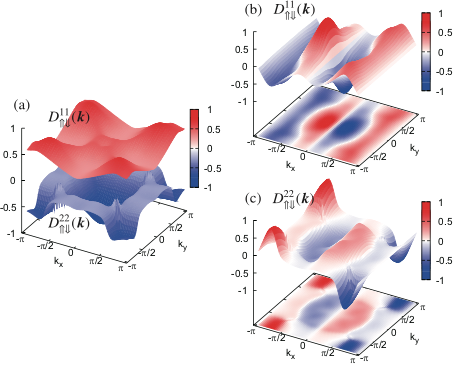} 
\caption{(color online) 
SC gap functions at the $m=1$ and $2$ energy bands for (a) the pseudospin singlet pairing and (b, c) the pseudospin triplet pairing, 
clearly showing $s_{\pm}$-wave symmetry and $p_{\pm}$-wave symmetry. 
We set $\lambda = 0.3$ eV and $n=4.7$, and the other parameters are the same as in Fig.~\ref{fig:EV_st}. 
} 
\label{fig:gap2}
\end{center} 
\end{figure}

On the other hand, as shown in Fig.~\ref{fig:gap2}, we find that 
the dominant SC gap functions have $s_{\pm}$-wave symmetry for the pseudospin singlet pairing and $p_\pm$-wave symmetry for the pseudospin triplet pairing 
in the parameter region shown in Fig.~\ref{fig:EV_st}. 
These pairing are formed in the $m=1$ and $m=2$ energy bands that have the strong character of $j_{\rm eff}=1/2$ and $j_{\rm eff}=3/2$, respectively (see Fig.~\ref{fig:jeff}). 
Notice also in Fig.~\ref{fig:gap2} that the SC gap functions alter the sign between $m=1$ and $m=2$. 
Therefore, the symmetry of the SC gap functions is $s_{\pm}$-wave or $p_\pm$-wave, instead of simply $s$-wave or $p$-wave.  
These pairings are favored for small $\lambda$ where the two band character is enhanced at the Fermi level 
by forming the large electron FS and the relatively small hole FS in the noninteracting limit (see Fig.~\ref{fig:En_DOS_FS}). 


\subsection{Pseudospin Pairing} 

To better understand the internal structure of the pseudospin pairing, 
let us decompose the pseudospin singlet pairing found in Figs.~\ref{fig:EV_dt}(a) and \ref{fig:EV_st}(a). 
The dominant components of the numerically obtained SC gap function $\hat{D}(\bm{k})$ 
can be decomposed into a mixture of the spin singlet and triplet pairing 
when it is represented in the $t_{2g}$ orbital basis with $\Delta_{\alpha \beta} (\bm{k})= \Delta^{ab}_{\sigma \sigma^{\prime}}(\bm{k})$. 
As shown in Table~\ref{tab:gap},  
the intra-orbital elements $\Delta^{aa}_{\sigma \bar{\sigma}}(\bm{k})$ $(a=yz, zx, xy)$ are spin singlet, 
i.e., $\Delta^{aa}_{\sigma \bar{\sigma}}(\bm{k}) = -\Delta^{aa}_{\bar{\sigma} \sigma}(\bm{k})$, 
while the inter-orbital ones are spin triplet, i.e, $\Delta^{ab}_{\sigma \sigma'}(\bm{k}) = \Delta^{ab}_{\sigma' \sigma}(\bm{k})$ for $a\ne b$, 
but the SC gap function changes the sign when the $t_{2g}$ orbital indices are exchanged, i.e., 
$\Delta^{yz \, zx}_{\sigma \bar{\sigma}}(\bm{k}) = -\Delta^{zx \, yz}_{\sigma \bar{\sigma}}(\bm{k})$, 
$\Delta^{yz \, xy}_{\sigma \sigma}(\bm{k}) = -\Delta^{xy \, yz}_{\sigma \sigma}(\bm{k})$, and 
$\Delta^{zx \, xy}_{\sigma \sigma}(\bm{k}) = -\Delta^{xy \, zx}_{\sigma \sigma}(\bm{k})$ 
for the dominant components. 
Since the spin SU(2) symmetry of the Hamiltonian is broken due to the presence of the SOC, 
the minus sign of the SC gap function $\hat{D}(\bm{k})$ for the psuedospin singlet pairing in Eq.~(\ref{eq:SCG}) 
can be imposed on either orbital or spin degree of freedom in $\Delta_{\alpha\beta}(\bm{k})= \Delta^{ab}_{\sigma \sigma^{\prime}}(\bm{k})$ represented 
in terms of the $t_{2g}$ basis to satisfy $\hat{\Delta}(-\bm{k}) = - \left[\hat{\Delta}(\bm{k})\right]^{\rm T}$. 

\begin{table} 
\begin{center} 
\begin{tabular}{|l|c|c|c|c|} \hline 
\multicolumn{5}{|c|}{ Pseudospin singlet: 
       $D^{mm'}_{\eta \eta^{\prime}}(\bm{k})= -D^{m'm}_{\eta^{\prime} \eta}(-\bm{k})$} \\ \hline 
$\Delta^{ab}_{\sigma \sigma^{\prime}}(\bm{k})$ & Parity ($\bm{k} \leftrightarrow -\bm{k}$) 
                                               & Orbital ($a \leftrightarrow b$) 
                                               & Spin ($\sigma \leftrightarrow \sigma^{\prime}$) 
                                               & $\Delta^{ba}_{\sigma^\prime \sigma}(-\bm{k})$ \\ \hline 
$\Delta^{xy \, xy}_{\sigma \bar{\sigma}}(\bm{k})$                                           & + & + & - &-  \\ \hline 
$\Delta^{yz \, yz}_{\sigma \bar{\sigma}} (\bm{k})$ & + & + & - &-   \\ \hline 
$ \Delta^{zx \, zx}_{\sigma \bar{\sigma}}(\bm{k})$ & + & + & -  &- \\ \hline 
$\Delta^{yz \, xy}_{\sigma \sigma}(\bm{k})$              & + & - & + &-  \\ \hline 
$ \Delta^{zx \, xy}_{\sigma \sigma}(\bm{k})$              & + & - & + &-  \\ \hline 
$\Delta^{yz \, zx}_{\sigma \bar{\sigma}}(\bm{k})$                                           & + & - & + &-  \\ \hline 
\end{tabular} 
\caption{
Dominant components (left column) of the SC gap function $\Delta^{ab}_{\sigma \sigma^{\prime}} (\bm{k})$ in the $t_{2g}$ basis
found in the numerically obtained pseudospin singlet SC gap function $D^{mm'}_{\eta \eta^{\prime}}(\bm{k})$ 
and the sign change of  $\Delta^{ab}_{\sigma \sigma^{\prime}} (\bm{k})$ 
under the parity transformation, orbital index exchange, and spin index exchange (three columns in the middle). 
The sign change of $\Delta^{ab}_{\sigma \sigma^{\prime}} (\bm{k})$ when it is transformed to $\Delta^{ba}_{\sigma^\prime \sigma}(-\bm{k})$ is also shown in the right column. 
} 
\label{tab:gap} 
\end{center} 
\end{table} 

The dominant components of the SC gap function $\Delta^{ab}_{\sigma \sigma^{\prime}} (\bm{k})$ 
found in the numerically obtained pseudospin singlet SC gap function $D^{mm'}_{\eta \eta^{\prime}}(\bm{k})$ is listed in Table~\ref{tab:gap}. 
Here, $m$ ($m'$) and $\eta$ ($\eta'$) are the band and pseudospin indices, respectively, for the energy band basis that diagonalizes 
the noninteracting Hamiltonian $H_0$ in Eq.~(\ref{eq:xi}), as defined in Eq.~(\ref{eq:band}). 
To understand these dominant components, 
let us consider the $d_{x^{2}-y^{2}}$-wave pseudospin singlet pairing formed by the energy band $m=1$ in the large SOC limit, as a simple example. 
As described in Sec.~\ref{sec:nonint}, the energy band basis is smoothly connected to the local atomic $j_{\rm eff}$ basis in the large SOC limit. 
Therefore, in this limit,  the $d_{x^{2}-y^{2}}$-wave pseudospin singlet pairing is formed by $a^{1}_{\bm{k}\Uparrow}$ and $a^{1}_{-\bm{k}\Downarrow}$, 
where the index ``1" corresponds to $j_{\rm eff}=1/2$ [see Eq.~(\ref{eq:jeff})]. 
Since the local atomic $j_{\rm eff}$ basis and the $t_{2g}$ orbital basis are transformed via the unitary matrix $\mathcal{W}$ in Eq.~(\ref{eq:uw}), 
the $d_{x^{2}-y^{2}}$-wave pseudospin singlet pairing in the large SOC limit is represented as 
\begin{equation}
a^{1}_{\bm{k}\Uparrow} a^{1}_{-\bm{k}\Downarrow} 
=  \frac{1}{3} 
   (  c^{yz}_{ \bm{k}\downarrow} -ic^{zx}_{ \bm{k}\downarrow} +c^{xy}_{ \bm{k}  \uparrow} ) 
   (  c^{yz}_{-\bm{k}  \uparrow} +ic^{zx}_{-\bm{k}  \uparrow} -c^{xy}_{-\bm{k}\downarrow} ).
\end{equation}
We can now observe that 
the $d_{x^{2}-y^{2}}$-wave pseudospin singlet pairing can be expressed as 
a linear combination of the spin singlet and triplet pairings formed by the $t_{2g}$ orbitals 
and these components correspond exactly to those of $\Delta^{ab}_{\sigma \sigma^{\prime}} (\bm{k})$ listed in Table~\ref{tab:gap}. 
In other words, this is a consequence of the fact that the pseudospin singlet superconductivity found in our calculations 
is formed essentially by the pseudospin $j_{\rm eff}=1/2$ quasiparticles. 

\subsection{$d_{x^{2}-y^{2}}$-wave pseudospin singlet pairing} 


In order to understand the pairing mechanism of the pseudospin singlet pairing, 
here we calculate, in analogy to the intrinsic spin susceptibility, the the longitudinal (out-of-plane) pseudospin susceptibility 
\begin{equation} 
\chi^{mn}_{J^{zz}_{\text{eff}}}(q) 
=  \frac{1}{N} \int_{0}^{\beta} d\tau \ \mathrm{e}^{\mathrm{i} \epsilon_{m} \tau} 
    \langle T_{\tau} J^{z \, m}_{\text{eff} \, \bm{q}}(\tau) J^{z \, n}_{\text{eff} \, -\bm{q}}(0) \rangle_{\text{c}} 
\end{equation} 
and the transverse (in-plane) pseudospin susceptibility 
\begin{equation}
\chi^{mn}_{J^{\pm}_{\text{eff}}}(q) 
=  \frac{1}{N} \int_{0}^{\beta} d\tau \ \mathrm{e}^{\mathrm{i} \epsilon_{m} \tau} 
    \langle T_{\tau} J^{- \, m}_{\text{eff} \, \bm{q}}(\tau) J^{+ \, n}_{\text{eff} \, -\bm{q}}(0) \rangle_{\text{c}} , 
\end{equation}
where the $z$-component of the pseudospin operator $J^{z \, m}_{\text{eff} \, \bm{q}}$ is defined as 
\begin{equation}
J^{z \, m}_{\text{eff} \, \bm{q}} 
=  \frac{1}{2} \sum_{\bm{k}} 
   \left(  a^{m\dagger}_{\bm{k} \Uparrow}   a^{m}_{\bm{k}+\bm{q} \Uparrow} 
          -a^{m\dagger}_{\bm{k} \Downarrow} a^{m}_{\bm{k}+\bm{q} \Downarrow} \right)  
\end{equation} 
and the correspond transverse components $J^{\pm \, m}_{\text{eff} \, \bm{q}}$ are defined as 
\begin{equation} 
J^{+ \, m}_{\text{eff} \, \bm{q}} = \sum_{\bm{k}} a^{m\dagger}_{\bm{k} \Uparrow}   a^{m}_{\bm{k}+\bm{q} \Downarrow}   
\end{equation} 
and 
\begin{equation} 
J^{- \, m}_{\text{eff} \, \bm{q}} = \sum_{\bm{k}} a^{m\dagger}_{\bm{k} \Downarrow} a^{m}_{\bm{k}+\bm{q} \Uparrow}   
\end{equation} 
with $\left[ J^{- \, m}_{\text{eff} \, \bm{q}} \right]^{\dagger} = J^{+ \, m}_{\text{eff} \, -\bm{q}}$. 
Here, $a^{m\dagger}_{\bm{k} \eta}$ ($a^m_{\bm{k} \eta}$) is the Fourier transform of the $j_{\rm eff}$ operator $a^{m\dagger}_{i \eta}$ ($a^m_{i \eta}$) defined in Eq.~(\ref{eq:jeff}). 

\begin{figure}
\begin{center}
\includegraphics[clip, width=8.0cm]{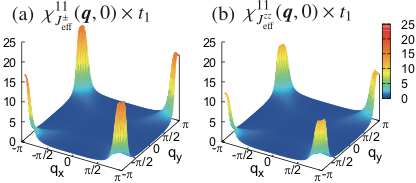}  
\caption{(color online) 
(a) Transverse and (b) longitudinal pseudospin susceptibilities in the $j_{\text{eff}}=1/2$ channel, 
$\chi^{11}_{J^{\pm}_{\text{eff}}}(\bm{q}, i\epsilon_{m}=0)$ and $\chi^{11}_{J^{zz}_{\text{eff}}}(\bm{q}, i\epsilon_{m}=0)$, respectively. 
The parameters are $\lambda =0.5$ eV, $U=0.75$ eV, $J/U=0.05$, and $n=5$. 
The other parameters are given in Eq.~(\ref{eq:parameter}). 
} 
\label{fig:Jpm_zz} 
\end{center} 
\end{figure}

Figure~\ref{fig:Jpm_zz} shows the pseudospin susceptibilities for $\lambda =0.5$ eV, $U=0.75$ eV, $J/U=0.05$, and $n$=5. 
Note that the susceptibilities are multiplied by $t_{1}$ to make them dimensionless. 
Although the transverse and longitudinal pseudospin susceptibilities are quantitatively different  due to a finite Hund's coupling,~\cite{Jackeli09}
they both exhibit peak structures around momentum $\bm{Q}=(\pi, \pi)$ and other symmetrically equivalent momenta, 
and thus the AF pseudospin instability is enhanced for this set of parameters. 
This instability can be understood by the FS topology shown in Fig.~\ref{fig:En_DOS_FS}(a):  
The shape of the FS satisfies nearly the perfect nesting condition with the nesting vector $\bm{Q}=(\pi, \pi)$ and other symmetrically equivalent momenta. 
In addition, we notice in Fig.~\ref{fig:Jpm_zz} that 
the transverse pseudospin susceptibility shows a larger peak value than the longitudinal one. 
This implies that the in-plane AF order is more favored than the out-of-plane one, 
which is consistent with the experimental observation for Sr$_2$IrO$_4$~\cite{Kim08, Kim09}. 

\begin{figure} 
\begin{center}
\includegraphics[clip, width=8.0cm]{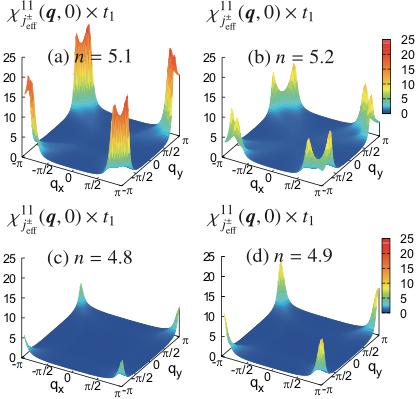}  
\caption{(color online)  
Transverse pseudospin susceptibility $\chi^{11}_{J^{\pm}_{\text{eff}}}(\bm{q}, i\epsilon_{m}=0)$ in the $j_{\text{eff}}=1/2$ channel 
for the electron doping at (a) $n=5.1$ and (b) $n=5.2$ and for the hole doping at (c) $n=4.8$ and (d) $n=4.9$. 
The other parameters are the same as in Fig.~\ref{fig:Jpm_zz}
} 
\label{fig:Jpms} 
\end{center} 
\end{figure}

Figure~\ref{fig:Jpms} displays the $n$ dependence of the transverse pseudospin susceptibility. 
As $n$ departs away from $n=5$, the peak values monotonically decrease 
and thus the pseudospin AF fluctuations are suppressed. Accordingly, the pairing interaction forming the Cooper pairs becomes weaker. 
We can also notice in Fig.~\ref{fig:Jpms} that 
the doping dependence of the pseudospin susceptibility from $n=5$ is qualitatively different for the electron-doped case ($n>5$) and the hole-doped case ($n<5$). 
The single peak appearing at $\bm{Q}=(\pi, \pi)$ for $n=5$ moves slightly along $x$ and $y$ directions to $(\pi,\pi\pm\delta)$ and $(\pi\pm\delta,\pi)$ for the electron doping $n>5$, 
while the single peak structure remains at $\bm{Q}=(\pi, \pi)$ for the hole doping $n<5$. 
Similar behavior of the spin susceptibility is found in the single-band Hubbard model for cuprate superconductors~\cite{Yanase03}. 
Note also that the hole pockets around $\bm{Q}=(\pi, \pi)$ and other symmetrically equivalent momenta appear for $n<4.9$ in the noninteracting limit (see Fig.~\ref{fig:lam_n}), 
which also reduces significantly $\chi^{11}_{j^{\pm}_{\text{eff}}}(\bm{q}, i\epsilon_{m}=0)$ and $\chi^{11}_{j^{zz}_{\text{eff}}}(\bm{q}, i\epsilon_{m}=0)$ 
since other scattering processes arise due to the appearance of the additional Fermi surface. 

From these results of the pseudospin susceptibilities, 
we can now understand the doping and $\lambda$ dependence of $\lambda_{\text{e}}$ in Fig.~\ref{fig:EV_dt}(a). 
For the electron doping with $n>5$, 
the DOS at the Fermi level increases toward the vHS around $n=5.2$--$5.3$ depending on $\lambda$ (see Fig.~\ref{fig:lam_n}), 
while the pseudospin AF instability decreases with departing away from $n=5$. 
As a result of this competition, $\lambda_{\text{e}}$ exhibits a maximum around $n \sim 5.1$, as shown in Fig.~\ref{fig:EV_dt}(a). 
Furthermore, the peak value of $\lambda_{\text{e}}$ is enhanced with increasing $\lambda$, especially for $\lambda > 0.4$ eV. 
This is because the hole pockets around $\bm{k}=(\pi, \pi)$ and symmetry equivalent momenta disappear for $\lambda > 0.4$ eV 
and only the single band crosses the Fermi level in the noninteracting limit (Fig.~\ref{fig:lam_n}). 
Therefore, it is understood that the $d_{x^{2}-y^{2}}$-wave pseudospin singlet pairing found here in our calculations is mediated by the AF pseudospin fluctuations. 
This is an analogy to cuprate superconductors where the $d_{x^{2}-y^{2}}$-wave spin $s=1/2$ singlet pairing is mediated by the AF spin fluctuations. 

\begin{figure} 
\begin{center}
\includegraphics[clip, width=8.cm]{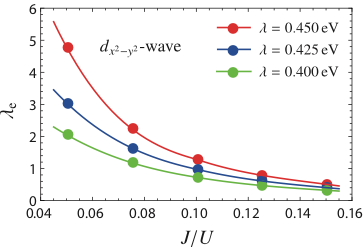}  
\caption{(color online)  
The largest eigenvalue $\lambda_{\text{e}}$ of the linearized Eliashberg equation vs. $J/U$ 
for the $d_{x^{2}-y^{2}}$-wave pseudospin singlet pairing in the electron doped case at $n=5.2$ with $U=0.8$ eV 
and three different values of $\lambda$ indicated in the figure. The other parameters are given in Eq.~(\ref{eq:parameter}). 
} 
\label{fig:EV_Jd} 
\end{center} 
\end{figure}

One of the important differences from the spin $s=1/2$ singlet pairing in cuprate superconductors is that 
the pseudospin singlet pairing found here is rather sensitive to the Hund's coupling $J$. 
Figure~\ref{fig:EV_Jd} shows 
the $J/U$ dependence of $\lambda_{\text{e}}$  for the electron doping at $n=5.2$ with several $\lambda$ values. 
As we have discussed above, 
the $d_{x^{2}-y^{2}}$-wave pseudospin $j_{\text{eff}}=1/2$ singlet pairing becomes weaker with decreasing $\lambda$. 
Moreover, it is also suppressed with increasing $J/U$. 
This is because of the suppression of the pseudospin AF fluctuations due to the decrease of the inter-orbital on-site 
interaction $U^{\prime}$ through $U^{\prime}=U-2J$. 

\begin{figure} 
\begin{center}
\includegraphics[clip, width=8.0cm]{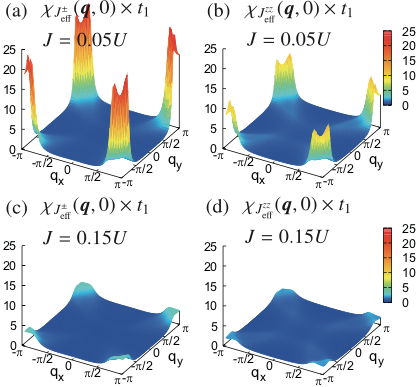}  
\caption{(color online)  
Total transverse and longitudinal pseudospin susceptibilities, 
$\chi_{J^{\pm}_{\text{eff}}}(\bm{q}, i\epsilon_{m}=0)$ and $\chi_{J^{zz}_{\text{eff}}}(\bm{q}, i\epsilon_{m}=0)$, respectively, 
for (a, b) $J/U=0.05$ and (c, d) $J/U=0.15$. 
The parameters are $\lambda =0.5$ eV, $U=0.75$ eV, and $n=5.1$. 
The other parameters are given in Eq.~(\ref{eq:parameter}). 
} 
\label{fig:XJpm_zz_J} 
\end{center} 
\end{figure}

Figure~\ref{fig:XJpm_zz_J} shows the longitudinal and transverse components, 
\begin{equation} 
\chi_{J^{zz}_{\text{eff}}}(q) = \sum_{m,n} \chi^{mn}_{J^{zz}_{\text{eff}}}(q) 
\end{equation}
and
\begin{equation}
\chi_{J^{\pm}_{\text{eff}}}(q) = \sum_{m,n} \chi^{mn}_{J^{\pm}_{\text{eff}}}(q) , 
\end{equation} 
respectively, of the total pseudospin susceptibility for small and large values of $J$, i.e., $J/U=0.05$ and $J/U=0.15$. 
It is clearly observed that the pseudospin susceptibilities are significantly suppressed with increasing $J/U$, 
and accordingly the $d_{x^{2}-y^{2}}$-wave pseudospin singlet pairing is also suppressed. 

\begin{figure}
\begin{center} 
\includegraphics[width=7.0cm,clip]{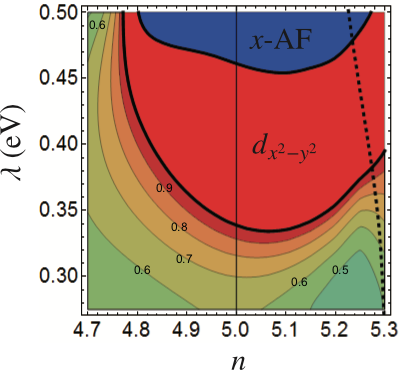} 
\caption{(color online) 
Contour plot of the largest eigenvalue $\lambda_{\text{e}}$ of the linearized Eliashberg equation for the $d_{x^{2}-y^{2}}$-wave pseudospin singlet pairing 
in the parameter space of the electron density $n$ and the SOC $\lambda$ for $U$=0.8 eV, and $J/U$=0.05. 
The other parameters are given in Eq.~(\ref{eq:parameter}). 
The red region represents the $d_{x^{2}-y^{2}}$-wave pseudospin singlet SC phase where $\lambda_{\text{e}}\ge1$. 
The $x$-AF phase is also indicated by blue where $\max[ \hat{V} \hat{\chi}_{0}(q) ]\ge1$. 
The black dashed line denotes the $(n,\lambda)$ region where the vHS appears exactly at the Fermi level 
in the noninteracting limit. 
} 
\label{fig:PD_d}
\end{center} 
\end{figure}

Finally, we systematically explore the stability of the $d_{x^{2}-y^{2}}$-wave pseudospin singlet pairing 
in the parameter space of the electron concentration $n$ and the SOC $\lambda$. 
The obtained phase diagram in Fig.~\ref{fig:PD_d} shows 
the contour plots of the largest eigenvalue of the linearized Eliashberg equation for the $d_{x^{2}-y^{2}}$-wave pseudospin singlet pairing 
in the parameter set of $U=0.8$~eV and $J/U=0.05$, 
where the $d_{x^{2}-y^{2}}$-wave pseudospin singlet pairing is favored over the $s_\pm$-wave pseudospin singlet pairing. 
The red colored region represents the $d_{x^{2}-y^{2}}$-wave pseudospin singlet SC phase 
and  the blue one is the $x$-AF phase where the AF spin moment is aligned within the $xy$-plane. 
The SC and AF phases here are identified when  $\lambda_{\text{e}}\ge1$ and $\max[ \hat{V} \hat{\chi}_{0}(q) ]\ge1$, respectively. 

\subsection{$s_{\pm}$-wave pseudospin singlet pairing} 

As we discussed in Sec.~\ref{sec:PS}, the $s_{\pm}$-wave pseudospin singlet pairing is more favored than the $d_{x^{2}-y^{2}}$-wave pseudospin singlet pairing 
when the single band character is less pronounced and the Hund's coupling $J/U$ is large.  
To understand the pairing mechanism of the $s_{\pm}$-wave pseudospin singlet pairing, 
let us consider the longitudinal and transverse spin susceptibilities,  
\begin{equation}
\chi^{ab}_{S^{zz}}(q) 
=  \frac{1}{N} \int_{0}^{\beta} d\tau \ \mathrm{e}^{i\epsilon_{m} \tau} 
   \langle T_{\tau} S^{z \, a}_{\bm{q}}(\tau) S^{z \, b}_{-\bm{q}}(0) \rangle_{\text{c}}
\end{equation} 
and 
\begin{equation}
\chi^{ab}_{S^{\pm}}(q) 
=  \frac{1}{N} \int_{0}^{\beta} d\tau \ \mathrm{e}^{i\epsilon_{m} \tau} 
   \langle T_{\tau} S^{- \, a}_{\bm{q}}(\tau) S^{+ \, b}_{-\bm{q}}(0) \rangle_{\text{c}},
\end{equation} 
where the longitudinal and transverse components of the spin operator, $S^{z \, a}_{ \bm{q}}$ and $S^{\pm \, a}_{ \bm{q}}$, for $t_{2g}$ orbital $a\,=(yz,\,zx,\,xy)$ are given as 
\begin{equation}
S^{z \, a}_{\bm{q}} 
=  \frac{1}{2} \sum_{\bm{k}} 
    \left(  c^{a\dagger}_{\bm{k} \uparrow}      c^{a}_{\bm{k}+\bm{q} \uparrow} 
            -c^{a\dagger}_{\bm{k} \downarrow} c^{a}_{\bm{k}+\bm{q} \downarrow} \right)  , 
\end{equation} 
\begin{equation}
S^{+ \, a}_{ \bm{q}} = \sum_{\bm{k}} c^{a\dagger}_{\bm{k} \uparrow}   c^{a}_{\bm{k}+\bm{q} \downarrow} , 
\end{equation}
and 
\begin{equation}
S^{- \, a}_{ \bm{q}} = \sum_{\bm{k}} c^{a\dagger}_{\bm{k} \downarrow}   c^{a}_{\bm{k}+\bm{q} \uparrow} 
\end{equation}
with $\left[ S^{- \, a}_{ \bm{q}} \right]^{\dagger} = S^{+ \, a}_{ -\bm{q}}$. 

\begin{figure} 
\begin{center}
\includegraphics[clip, width=8.5cm]{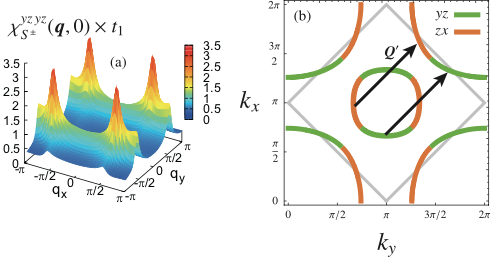} 
\caption{(color online). 
(a) Transverse spin susceptibility in the $yz$ orbital $\chi^{yz,yz}_{S^{\pm}}(\bm{q},\mathrm{i} \epsilon_{m}=0)$ for $n=4.7$, $\lambda = 0.3$ eV, $U=0.9$ eV, and $J/U=0.3$. The other parameters are given in Eq.~(\ref{eq:parameter}). 
(b) Noninteracting Fermi surfaces decomposed with the $yz$- and $zx$-orbital components. 
The parameters correspond to (a). The nesting vector $\bm{Q}^{\prime}$ is indicated. 
} 
\label{fig:spm} 
\end{center} 
\end{figure}

Figure~\ref{fig:spm}(a) shows the transverse spin susceptibility in the $yz$ orbital $\chi^{yz \, yz}_{S^{\pm}}(\bm{q},\mathrm{i} \epsilon_{m} = 0)$ 
for $n=4.7$ with $\lambda = 0.3$ eV, $U=0.9$ eV, and $J/U=0.3$. 
Note that the spin susceptibility in the $zx$ orbital $\chi^{zx \, zx}_{S^{\pm}}(\bm{q},\mathrm{i} \epsilon_{m} = 0)$ 
is obtained from $\chi^{yz \, yz}_{S^{\pm}}(\bm{q},\mathrm{i} \epsilon_{m} = 0)$ simply by rotating the $yz$-orbital by 90 degrees. 
We can clearly observe in Fig.~\ref{fig:spm}(a) that 
$\chi^{yz \, yz}_{S^{\pm}}(\bm{q},\mathrm{i} \epsilon_{m} = 0)$ exhibits peaks at $\bm{Q}^{\prime} \sim (5\pi/8, 5\pi/8)$ and symmetrically equivalent momenta. 
Note however that 
the spin susceptibility in the $xy$-orbital $\chi^{xy \, xy}_{S^{\pm}}(\bm{q},\mathrm{i} \epsilon_{m} = 0)$ is found small 
compared to those in the $yz$ and $yz$ orbitals simply 
because the $xy$-orbital energy level is located lower than the $yz$- and $yz$-orbital energy levels and thus it is almost complete occupied by electrons. 
As shown in Fig.~\ref{fig:spm}(b), 
this momentum $\bm{Q}^{\prime}$ corresponds to the nesting vector connecting the electron-like and hole-like FSs with the large component of the $yz$ orbital character. 
This nesting feature is exactly the same for the $zx$ orbital [see Fig.~\ref{fig:spm}(b)], 
which thus causes the same four peak structures in $\chi^{zx \, zx}_{S^{\pm}}(\bm{q},\mathrm{i} \epsilon_{m} = 0)$. 
Similarly, the longitudinal spin susceptibilities in the $yz$ and $zx$ orbitals exhibits four peaks at $\bm{Q}^{\prime}$ and symmetry equivalent momenta. 
Furthermore, we find that 
the nesting vector $\bm{Q}^{\prime}$ tends to change the sigh of the SC gap functions 
because the inter-band scattering through the nesting vector $\bm{Q}^{\prime}$ is repulsive 
and thus the sign of the SC gap functions in the energy bands $m=1$ and $2$ is opposite. 
Therefore, the spin fluctuations with the nesting vector $\bm{Q}^{\prime}$ induce the $s_{\pm}$-wave pseudospin singlet pairing. 
This is analogous to the pairing mechanism for the iron-based superconductors: 
they also possess multi-band FSs and the inter-band repulsive scattering is expected to be dominant, 
which favors the $s_{\pm}$-wave intrinsic spin singlet pairing.~\cite{Kuroki08, Kuroki09} 


\begin{figure} 
\begin{center}
\includegraphics[clip, width=8.cm]{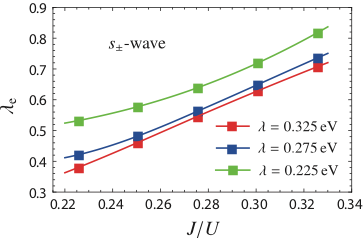}  
\caption{(color online)  
The largest eigenvalue $\lambda_{\text{e}}$ of the linearized Eliashberg equation vs. $J/U$ 
for the $s_{\pm}$-wave pseudospin singlet pairing in the hole doped region at $n=4.7$ 
with $U=0.9$ eV and three different values of $\lambda$ indicated in the figure. 
The other parameters are given in Eq.~(\ref{eq:parameter}). 
} 
\label{fig:EV_Js} 
\end{center} 
\end{figure}

It is now apparent that the two band feature is essential for stabilizing the $s_{\pm}$-wave pseudospin singlet pairing, 
which explains why the small $\lambda$ value is favored for this singlet pairing. 
In addition, we find that 
the $s_{\pm}$-wave pseudospin singlet pairing is more favored when the Hund's coupling $J$ is large. 
Figure~\ref{fig:EV_Js} shows the $J/U$ dependence of the eigenvalue $\lambda_{\text{e}}$ of the linearized Eliashberg equation in the hole doping at $n=4.7$ for several $\lambda$ values. 
We can clearly observe that 
the $s_{\pm}$-wave pseudospin singlet pairing is enhanced with increasing $J/U$ as well as with decreasing $\lambda$. 
This is because the spin fluctuations with the nesting vector $\bm{Q}^{\prime}$ is enhanced as $J/U$ is increased. 

\begin{figure} 
\begin{center}
\includegraphics[clip, width=8.0cm]{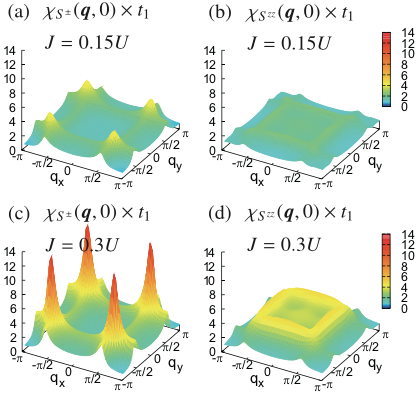}  
\caption{(color online)  
Total transverse and longitudinal spin susceptibilities, 
$\chi_{S^{\pm}}(\bm{q}, i\epsilon_{m}=0)$ and $\chi_{S^{zz}}(\bm{q}, i\epsilon_{m}=0)$, respectively, 
for (a, b) $J/U=0.15$ and (c, d) $J/U=0.3$. 
The parameters are $\lambda =0.3$ eV, $U=0.9$ eV, and $n=4.7$. 
The other parameters are given in Eq.~(\ref{eq:parameter}). 
} 
\label{fig:XSpm_zz_J} 
\end{center} 
\end{figure}

Figure~\ref{fig:XSpm_zz_J} shows the longitudinal and transverse components of the total spin susceptibility, 
\begin{equation} 
\chi_{S^{zz}}(q) = \sum_{a,b} \chi^{ab}_{S^{zz}}(q) 
\end{equation}
and
\begin{equation}
\chi_{S^{\pm}}(q) = \sum_{a,b} \chi^{ab}_{S^{\pm}}(q) , 
\end{equation} 
respectively, for small and large values of $J$, i.e., $J/U=0.15$ and $J/U=0.3$. 
It is indeed clearly observed that the total spin susceptibilities are significantly enhanced with increasing $J/U$. 
Accordingly, the $s_{\pm}$-wave pseudospin singlet pairing is also enhanced. 

\begin{figure}
\begin{center} 
\includegraphics[width=7.0cm,clip]{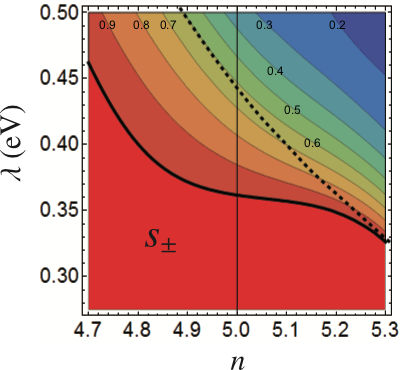} 
\caption{(color online) 
Contour plots of largest eigenvalue $\lambda_{\text{e}}$ of the linearized Eliashberg equation for the 
$s_{\pm}$-wave pseudospin singlet pairing in the parameter space of 
the electron density $n$ and the SOC $\lambda$ for $U=0.9$ eV and $J/U=0.3$. 
The other parameters are given in Eq.~(\ref{eq:parameter}). 
The red region represents the $s_{\pm}$-wave pseudospin singlet SC phase where $\lambda_{\text{e}}\ge1$. 
In the region above (below) the black dashed line, there exists one (two) FS(s) in the noninteracting limit 
(see also Fig.~\ref{fig:lam_n}). 
} 
\label{fig:PD_s}
\end{center} 
\end{figure}

Finally, we systematically explore the stability of the $s_{\pm}$-wave pseudospin singlet pairing 
in the parameter space of the electron concentration $n$ and the SOC $\lambda$. 
The obtained phase diagram in Fig.~\ref{fig:PD_s} shows the contour plots of the largest eigenvalue 
of the linearized Eliashberg equation 
for the $s_{\pm}$-wave pseudospin singlet pairing in the parameter set of 
$U=0.9$ eV and $J/U=0.3$, where the $s_{\pm}$-wave pseudospin singlet pairing is favored over 
the $d_{x^{2}-y^{2}}$-wave pseudospin singlet pairing. 
The red colored region represents the $s_{\pm}$-wave pseudospin singlet SC phase where $\lambda_{\text{e}}\ge1$. 

\section{Summary and Discussions}\label{sec:summary}

To summarize, we have studied possible superconductivity in carrier doped 5$d$ transition metal oxide Sr$_{2}$IrO$_{4}$ 
based on an effective $t_{2g}$ three-orbital Hubbard model on the square lattice with a large SOC. 
The linearized Eliashberg equation for the SC gap function was solved numerically with the RPA. 
We revealed that the $d_{x^{2}-y^{2}}$-wave pseudospin singlet pairing is expected to appear in the realistic parameter region, 
i.e., large SOC $\lambda$ and small Hund's coupling $J/U$. 
The $d_{x^{2}-y^{2}}$-wave pseudospin singlet pairing is induced by the AF pseudospin fluctuations 
and is favored for large SOC especially in the electron doped region. 
Furthermore, we have found that the 
$s_{\pm}$-wave pseudospin singlet pairing is stabilized for small $\lambda$ and large $J/U$, 
and is  more favored in the hole doped region. 
The $s_{\pm}$-wave pseudospin singlet pairing is induced by the inter band scattering 
arising from the intra-orbital AF spin fluctuations.  

The $d_{x^{2}-y^{2}}$-wave pseudospin singlet pairing found in large $\lambda$ and small $J/U$ agrees with the previous studies~\cite{Watanabe13, Yang14, Meng14}. 
In this region, the single-band nature with strong character of pseudospin $j_{\text{eff}}=1/2$ is enhanced especially in the electron doped region. 
Therefore, the $d_{x^{2}-y^{2}}$-wave pseudospin singlet pairing found here  is an analogy to the $d_{x^{2}-y^{2}}$-wave spin $s=1/2$ singlet pairing in cuprate superconductors. 
The realistic parameter for Sr$_{2}$IrO$_{4}$ is in this parameter region 
because it is estimated that  $\lambda \sim 0.37 \ \text{eV}$ and $J/U \sim 0.07$~\cite{Watanabe10, Arita12, Watanabe13}. 
Therefore, the $d_{x^{2}-y^{2}}$-wave pseudospin singlet pairing in the electron doped region is the most promising candidate for possible superconductivity in Sr$_{2}$IrO$_{4}$. 

The feature found here is similar to the case of cuprate superconductors from the viewpoint of purification of 
the $d_{x^{2}-y^{2}}$ orbital. 
H. Sakakibara \textit{et al.}~\cite{Sakakibara10, Sakakibara12, Sakakibara14} have suggested that 
SC critical temperature $T_{\text{c}}$ of cuprate superconductors tends to increase with reducing the hybridization of the $d_{3z^{2}-r^2}$ orbital from the $d_{x^{2}-y^{2}}$ one, 
thus purifying the $d_{x^{2}-y^{2}}$ orbital, 
which mainly contributes to the $d_{x^{2}-y^{2}}$-wave spin singlet pairing. 
In the same way, we found here that 
the largest eigenvalue of the linearized Eliashberg equation increases with increasing the SOC, 
purifying the band with the strong character of $j_{\text{eff}}=1/2$, which contributes most to the $d_{x^{2}-y^{2}}$-wave pseudospin singlet pairing. 

We found the $s_{\pm}$-wave pseudospin singlet pairing in the small $\lambda$ and large $J/U$ region, 
which is different from the results in the previous studies~\cite{Yang14, Meng14}. 
Z. Y. Meng \textit{et al.}~\cite{Meng14} have suggested a $p$-wave pairing in a hole doped region. 
However, we found that a pseudospin singlet pairing is always favored over a pseudospin triplet pairing.  
On the other hand, Y. Yang \textit{et al.}~\cite{Yang14} have proposed a $s_{\pm}$-wave pairing using the functional renormalization group method, 
and they have also suggested that ferromagnetic (FM) fluctuations are responsible for the pairing mechanism. 
FM fluctuations may be enhanced when the Fermi pocket is small. 
However, the enhancement cannot be found in our RPA calculations. 
Instead, our calculations show that the $s_{\pm}$-wave pairing is induced by the inter-band scattering arising from the intra-orbital spin fluctuations, without strong FM fluctuations. 

Finally, we briefly comment on $4d$ transition metal oxide Sr$_2$RhO$_4$. 
When the tight-binding parameters are estimated form the first-principles band structure calculations~\cite{Kim08},  
it is found that the ratios of the tight-binding parameters of the three $t_{2g}$ orbitals, 
$(t_{1},t_{2},t_{3},t_{4},t_{5},\mu_{xy})/t_{1}$, for Sr$_2$RhO$_4$ are almost the same as those for Sr$_2$IrO$_4$, 
except that the ratio of the SOC to $t_{1}$, $\lambda/t_{1} \sim 0.5$, is smaller than that for Sr$_2$IrO$_4$ ($\lambda/t_{1} \sim 1.0$). 
Therefore, based on our calculations here, we expect that 
the superconductivity with $s_{\pm}$-wave pairing symmetry would be favorable for Sr$_2$RhO$_4$ 
especially in the hole doped region, provided that  the Hund's coupling is sufficiently large.  

\section*{Acknowledgments} 

We would like to thank Toshihiro Sato, Kazuhiro Seki, Toshikaze Kariyado, and Hirofumi Sakakibara for useful discussions. 
The calculations were carried out using the supercomputer 
in Advanced Center for Computing and Communication centers of RIKEN and the HOKUSAI supercomputer at RIKEN. 
This work was supported in part by Grants-in-Aid for Scientific Research from JSPS (Projects No. 24740251 and No. 25287096) of Japan.
T.S. acknowledges Simons Foundation for financial support (award no. 534160). 



\end{document}